\documentclass[]{aa}
\usepackage{graphicx}
\usepackage{adjustbox}
\usepackage{txfonts}

\makeatletter
\renewcommand*\aa@pageof{, page \thepage{} of \pageref*{LastPage}}
\makeatother
\usepackage{hyperref}
\hypersetup{
    colorlinks=true,
    linkcolor=blue,
    filecolor=magenta,      
    urlcolor=blue,
    citecolor=blue,
    }
\defcitealias{marinfranch2009}{MF09}
\defcitealias{dotter2010}{D10}
\defcitealias{vandenberg2013}{V13}
\defcitealias{tailo2020}{T20}

\begin{document} 

\title{On the original composition of the gas forming first-generation stars in clusters: insights from HST and JWST}
\subtitle{}

\author{M. V. Legnardi\inst{1} \and A. P. Milone\inst{1,2} \and G. Cordoni\inst{3} \and A. F. Marino\inst{2,4} \and E. Dondoglio\inst{2} \and S. Jang\inst{5} \and E. P. Lagioia\inst{6} \and F. Muratore\inst{1} \and T. Ziliotto\inst{1} \and E. Bortolan\inst{1} \and A. Mohandasan\inst{1}
}

\institute{Dipartimento di Fisica e Astronomia ``Galileo Galilei'', Univ. di Padova, Vicolo dell'Osservatorio 3, Padova, IT-35122 \\ \email{mariavittoria.legnardi@phd.unipd.it}
\and
Istituto Nazionale di Astrofisica - Osservatorio Astronomico di Padova, Vicolo dell’Osservatorio 5, Padova, IT-35122
\and 
Research School of Astronomy and Astrophysics, Australian National University, Canberra, ACT 2611, Australia
\and
Istituto Nazionale di Astrofisica - Osservatorio Astrofisico di Arcetri, Largo Enrico Fermi, 5, Firenze, IT-50125
\and
Center for Galaxy Evolution Research and Department of Astronomy, Yonsei University, Seoul 03722, Korea
\and
South-Western Institute for Astronomy Research, Yunnan University, Kunming, 650500 P. R. China}

\titlerunning{The original composition of the gas forming the first stars in clusters} 
\authorrunning{Legnardi et al.}

\date{Received 8 February 2023 / Accepted 22 April 2023}

\abstract{Globular cluster (GC) stars composed of pristine material (first-generation, 1G, stars) are not chemically homogeneous, as they exhibit extended sequences in the "Chromosome Map" (ChM). Recent studies characterized 1G stars within the center of 55 Galactic GCs, revealing metallicity variations. Despite this progress, several unanswered questions persist, particularly concerning the link between the 1G metallicity spread and factors such as the radial distance from the cluster center or the host GC parameters. Additionally, it remains unclear whether the extended 1G sequence phenomenon is exclusive to old Galactic GCs with multiple populations. This work addresses these open issues, examining 1G stars in different environments. First, we combine Hubble Space Telescope (HST) and James Webb Space Telescope photometry of the GC 47\,Tucanae to study 1G stars at increasing distances from the cluster center. We find that metal-rich 1G stars are more centrally concentrated than metal-poor ones,  suggesting a metallicity radial gradient. Additionally, the two groups of 1G stars share similar kinematics. Since our analysis focuses on giant stars in the cluster center and M dwarfs in external fields, we discuss the possibility that the metallicity distribution depends on stellar mass. Subsequently, we analyze HST multi-band photometry of two simple-population clusters, NGC\,6791 and NGC\,1783, revealing elongated sequences in the ChM associated with metallicity variations. Finally, we investigate the 1G color distribution in 51 GCs, finding no connections with the host cluster parameters. These results shed light on the complex nature of 1G stars, providing insights into the GC formation environment.}

\maketitle
%
\section{Introduction}
\label{sec:intro}

The presence of multiple stellar populations in Galactic globular clusters (GCs) is now firmly established \citep[see][for recent reviews]{bastian2018, milone2022}. This phenomenon generated several questions concerning the origin of GCs, many of which are still to be answered.   

The "Chromosome Map" \citep[ChM;][]{milone2015} is a powerful tool for investigating stellar populations within GCs. It is a pseudo-two-color diagram extremely sensitive to the chemical composition of distinct stellar populations, thereby enhancing their separation. The standard  $\Delta_{\rm {\it C}\,F275W, F336W, F438W}$ vs.\,$\Delta_{\rm F275W, F814W}$ ChM \citep{milone2017} is derived from photometry in the F275W, F336W, F438W, and F814W bands of the Hubble Space Telescope (HST). This photometric diagram is particularly useful to separate stellar populations within the red-giant branch (RGB) or the upper main sequence (MS). The advent of the James Webb Space Telescope (JWST) allowed the construction of new ChMs, derived by combining optical and near-infrared bands of the HST and the JWST. As an example, recent works \citep{milone2023a, ziliotto2023} introduced the $\Delta_{\rm F115W,F322W2}$ vs.\,$\Delta_{\rm F606W,F115W}$ and $\Delta_{\rm F090W,F277W}$ vs.\,$\Delta_{\rm F090W,F150W}$ pseudo-two color diagrams, optimized to separate distinct stellar populations in the M-dwarfs regime at the bottom of the MS. Despite their differences, all ChMs exhibit two main groups of stars, called first- (1G) and second-generation (2G). 1G stars are composed of pristine material reflecting the chemical composition of their natal cloud, whereas 2G members exhibit various degrees of chemical enrichment.      

In addition to their effectiveness in identifying and characterizing multiple stellar populations, ChMs also provide direct insights into the formation environment of GCs. In this context, a notable discovery based on RGB ChMs indicates that the 1G sequences in most Galactic GCs do not conform to simple stellar populations. Instead, they exhibit either a spread or a bimodality in the ChM \citep[see Figs.~3-7 of][for a comprehensive collection of ChMs in 57 GCs]{milone2017}. This phenomenon is not limited to RGB stars, as \cite{dondoglio2021} observed extended 1G sequences along the red horizontal branch (HB) in 12 clusters. More recently, the 1G color extension has been detected even in the ChMs of unevolved MS stars \citep{legnardi2022, milone2023a}, suggesting that chemical variations are likely imprinted in the environment where 1G stars originated.

In the standard ChM, 1G stars exhibit a significant dispersion in the direction of the x-axis while maintaining a relatively constant position along the y-axis. This observation implies a connection between the extension of 1G stars and variations in the effective temperature among stars with similar F814W magnitude. The spread in helium content can impact the effective temperature of stars, as initially observed by \cite{milone2015} to interpret the extended 1G sequence that they observed in the ChM of NGC\,2808. To account for this phenomenon, \cite{milone2015, milone2018b} explored various physical mechanisms and found that none of them could produce pure helium enrichment without concurrent changes in light elements. Alternatively, helium inhomogeneities may come from events that occurred in regions of the Universe where the baryon-to-photon ratio was significantly enhanced \citep[e.g.,][]{arbey2020}.

The spectroscopic investigation by \cite{marino2019a} suggested that variations in metallicity, rather than helium content, may be responsible for the extended color range observed in 1G stars. In their study, Marino and collaborators analyzed 18 1G stars within the Galactic GC NGC\,3201 and observed a dispersion in [Fe/H] of $\sim 0.1$ dex. Recently, the photometric research by \citet[][]{legnardi2022} further supported this conclusion. To distinguish between helium and metallicity effects, Legnardi and collaborators introduced a pseudo-two-magnitude diagram aimed at maximizing the separation between stellar populations with varying iron content. Their analysis revealed a spread of [Fe/H], rather than helium (Y), among 1G stars in the two investigated targets, namely NGC\,6362 and NGC\,6838. Expanding this finding to a large sample of GCs, \cite{legnardi2022} estimated that internal variations in iron content within the 1G of 55 Galactic GCs range from less than $\sim 0.05$ dex to $\sim 0.30$ dex and mildly correlate with GC mass. 

As discussed by \cite{marino2019a}, unresolved binaries among 1G stars can provide extended sequences in the ChM and spurious lower metallicity abundances for stars with low $\Delta_{\rm F275W,F814W}$ values. However, they showed that very large fractions of binaries are needed to reproduce the observations, in contrast with what is observed in Galactic GCs and intermediate-age Magellanic Cloud (MC) star clusters \citep[][]{milone2012a, mohandasan2024a}. Hence, they concluded that very likely binaries provide a minor contribution to the color extension of the 1G \citep[see also][]{kamann2020a, martins2021a}.
  
The discovery of chemical abundance variations among 1G stars provides the opportunity to constrain the formation mechanisms of the stellar populations in GCs \citep[e.g.,][]{dantona2016a}. The observed [Fe/H] variations could be indicative of either chemical inhomogeneities within the original pristine material where the 1G population formed or the result of internal stellar feedback processes taking place within the clusters themselves \citep{mckenzie2021}. Moreover, the evidence that 2G stars of NGC\,6362 and NGC\,6838 exhibit smaller metallicity variations than the 1G is consistent with the scenarios where 2G stars formed in a high-density environment in the cluster center \citep{legnardi2022}. Finally, the color extension of 1G stars in the ChM would constrain the late stages of stellar evolution in star clusters. Indeed, the 1G stars of GCs with the same metallicity but different HB morphology span different ranges of [Fe/H] \citep{legnardi2022}.

Despite extensive observational efforts, several aspects of the extended 1G sequence phenomenon remain unexplored. For instance, it is yet uncertain whether the metallicity distribution of 1G stars depends on factors such as the radial distance from the cluster center or other parameters associated with the host GCs. Similarly, the status of extended 1G sequences as a unique characteristic exclusive to Galactic GCs with multiple populations, or their potential existence in Galactic open clusters or massive intermediate-age star clusters in the MCs, remains unknown. This work aims to shed light on these points, thus providing a comprehensive analysis of chemical inhomogeneities among 1G stars in Galactic and extragalactic stellar clusters.

First, we study metallicity variations among 1G stars of the Galactic GC NGC\,104 (47\,Tucanae). In this context, 47\,Tucanae is an ideal target. Indeed, the evidence that 2G stars are more centrally concentrated than the 1G \citep[e.g.,][]{milone2012b, cordero2014a,  dondoglio2021, lee2022a}, indicates that this GC retains information on the radial distribution of its multiple populations at formation. Moreover, a recent spectroscopic analysis conducted by \cite{marino2023} uncovered that 1G stars in 47\,Tucanae display variations in [Fe/H] of the order of $\sim 0.1$ dex, thus corroborating previous findings obtained through photometry. The second part of the present research investigates whether clusters with no evidence of multiple populations exhibit an extended ChM sequence comparable to that of 1G stars in Galactic GCs. To do that, we focus on  the Galactic open cluster NGC\,6791, and the Large Magellanic Cloud (LMC) cluster NGC\,1783. Finally, we come back to Galactic GCs looking for possible relations between the metallicity distribution of 1G stars and the main parameters of the host cluster.

The paper is organized as follows. Section~\ref{sec:data} presents the datasets and the techniques employed to reduce them. Section~\ref{sec:47Tuc} is dedicated to 1G stars of 47\,Tucanae, whereas the two simple-population clusters investigated in this work and the Galactic GCs are the targets of Sect.~\ref{sec:SSPCs} and \ref{sec:GGCs}, respectively. Section~\ref{sec:concl} provides the summary of this work together with conclusions. 

\begin{figure}
    \centering
    \includegraphics[width=.9\columnwidth]{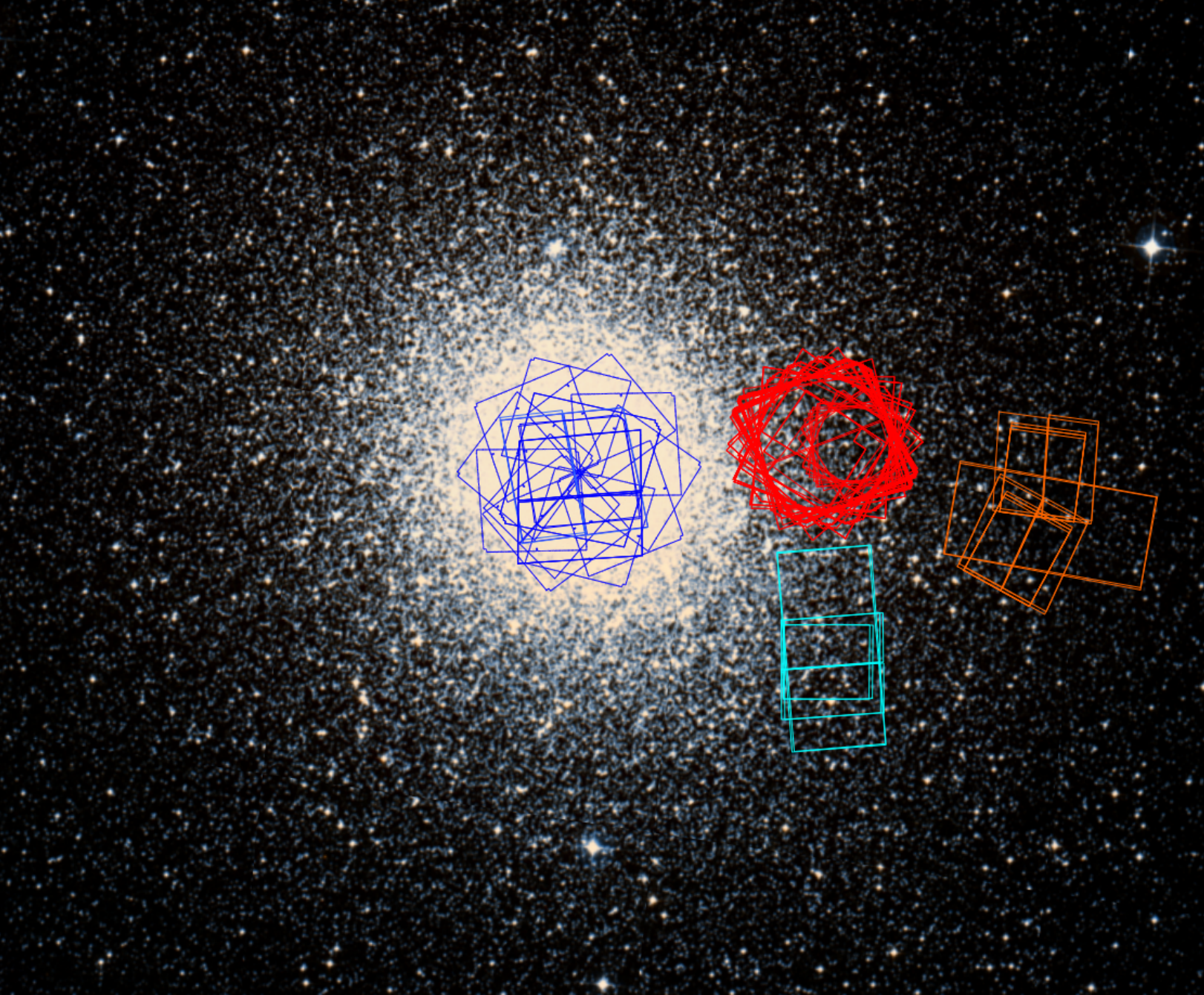}
    \caption{Footprints of the regions of 47\,Tucanae analyzed in this paper. The central region (blue) and field A (red) include only HST data, whereas field B (cyan) and C (orange) cover also JWST images.}
    \label{fig:map}
\end{figure}

\begin{figure}
    \centering
    \includegraphics[width=.95\columnwidth]{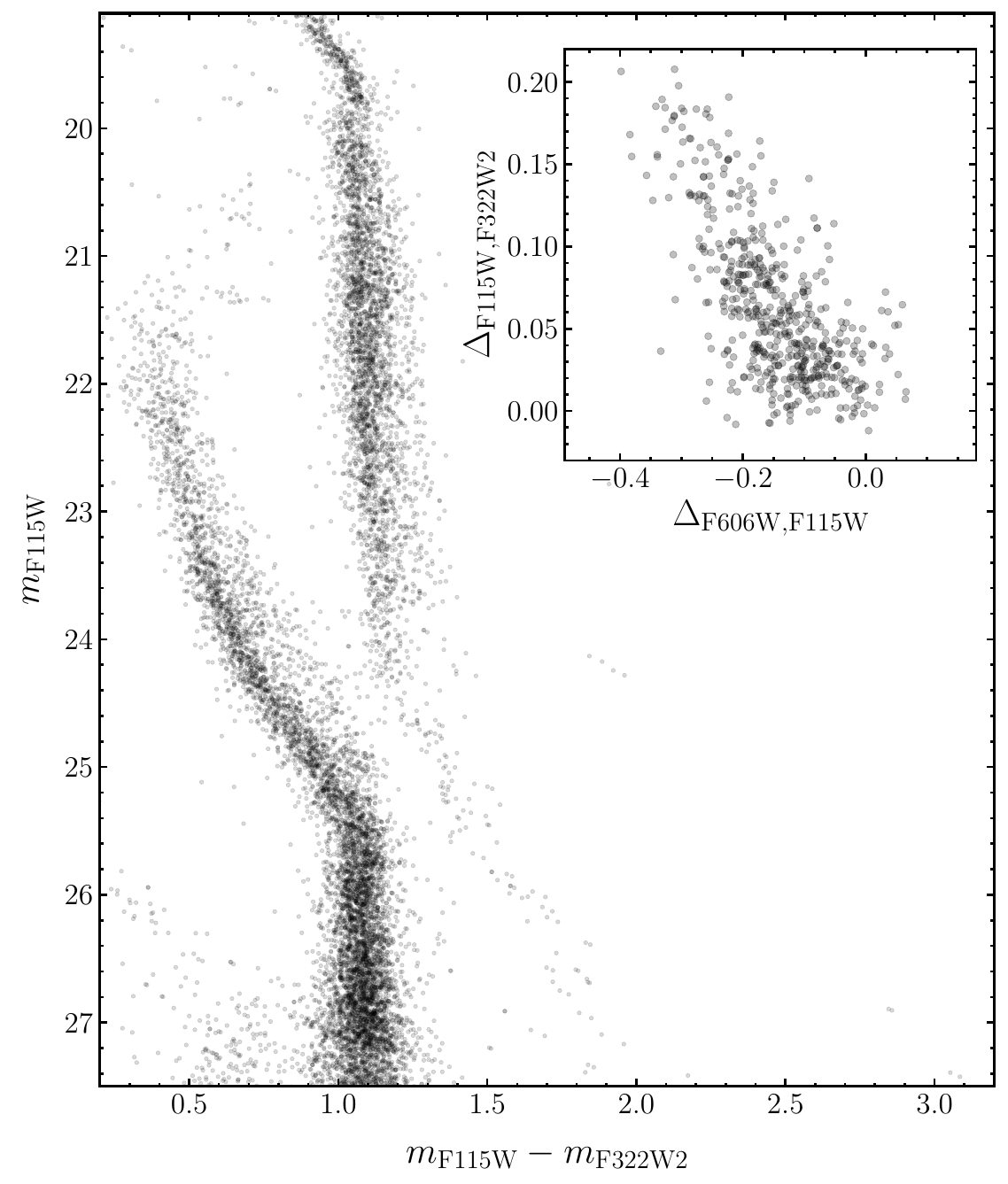}       
    \caption{$m_{\rm F115W}$ vs.\,$m_{\rm F115W}-m_{\rm F322W2}$ CMD of 47\,Tucanae stars within field C \citep{marino2024a}. The inset shows the $\Delta_{\rm F115W,F322W2}$ vs.\,$\Delta_{\rm F606W,F115W}$ ChM.}
    \label{fig:47Tuc_cmd}
\end{figure}

\section{Observations and data reduction}
\label{sec:data}
In the following, we describe the observations and the photometric and astrometric catalogs employed for this work. For simplicity, we discuss the data on 47\,Tucanae, on the simple-population star clusters NGC\,6791 and NGC\,1783, and the other Galactic GCs, separately. 

\subsection{47 Tucanae}
\label{subsec:47tuc}
We analyzed the chemical composition and the internal kinematics of 1G stars in the Galactic GC 47\,Tucanae by using data of four distinct regions, located at progressively larger distances from the cluster center. Figure~\ref{fig:map} illustrates the footprints of the four regions discussed in this paper, including the cluster center and three external fields, denoted as A, B, and C, respectively. 

The {\it Central field} (RA$\sim 00^{\rm h}24^{\rm m}06^{\rm s}$, DEC$\sim -72^{\rm d}04^{\rm m}53^{\rm s}$; blue footprints) covers the innermost region of $\sim 2.7\times2.7$ arcmin$^{2}$ within the cluster. In this area, we used the catalog by \cite{milone2017}, which includes photometry in the F275W and F336W bands of the Ultraviolet and Visual Channel of the Wide Field Camera 3 (UVIS/WFC3) and in the F435W, F606W, and F814W filters of the Wide Field Channel of the Advanced Camera for Survey (WFC/ACS) on board the HST. To study the kinematics of 1G stars in the core of 47\,Tucanae, we took advantage of stellar proper motions from \cite{libralato2022}. 

{\it Fields A} (RA$\sim 00^{\rm h}24^{\rm m}37^{\rm s}$, DEC$\sim -72^{\rm d}04^{\rm m}06^{\rm s}$; red footprints) and {\it B} (RA$\sim 00^{\rm h}22^{\rm m}36^{\rm s}$, DEC$\sim -72^{\rm d}09^{\rm m}27^{\rm s}$; cyan footprints) are located at distances of $\sim 7$ and $\sim 8.5$ arcmin west from the cluster center, respectively. For those regions, we relied on the photometric catalogs and proper motions derived recently by \cite{milone2023a}. Observations of field A were taken with the near-infrared channel of the Wide Field Camera 3 (IR/WFC3) through the F110W and F160W bands and with the WFC/ACS in the F606W and F814W filters. For field B, instead, Milone and collaborators used exposures collected through the near-infrared camera (NIRCam) on board the JWST. This field has been observed also with the UVIS/WFC3 in the F606W filter and with the IR/WFC3 in the F110W and F160W bands. We refer to the aforementioned papers and their respective references for details on the data reduction and the techniques used to measure stellar proper motions.

Finally, {\it field C} (RA$\sim 00^{\rm h}21^{\rm m}16^{\rm s}$, DEC$\sim -72^{\rm d}06^{\rm m}16^{\rm s}$; orange footprints) is located at a distance of $\sim 11$ arcmin west with respect to the cluster center. To investigate 1G stars in this outer region we used data obtained with the NIRCam/JWST as part of the GO-2560 program (PI: A. F. Marino), which includes images acquired with the F115W and the F322W2 filters. Moreover, we complemented these observations with HST exposures collected through the UVIS/WFC3 F606W and the IR/WFC3 F110W and F160W bands. The techniques adopted to reduce these observations, as well as to compute stellar proper motions, are described in \citet{marino2024a}.

Figure~\ref{fig:47Tuc_cmd} illustrates the $m_{\rm F115W}$ vs.\,$m_{\rm F115W}-m_{\rm F322W2}$ color-magnitude diagram (CMD) of 47\,Tucanae derived by \citet{marino2024a} for stars within field C. In close analogy with \citet{milone2023a}, this photometric diagram has been used together with the $m_{\rm F115W}$ vs.\,$m_{\rm F606W}-m_{\rm F115W}$ CMD to derive the $\Delta_{\rm F115W,F322W2}$ vs.\,$\Delta_{\rm F606W,F115W}$ ChM, plotted in the inset of Fig.~\ref{fig:47Tuc_cmd}. Here, 1G stars lay in proximity of the origin, whereas 2G stars extend from ($\Delta_{\rm F606W,F115W},\Delta_{\rm F115W,F322W2})\sim(-0.1,0.05)$ to $\sim (-0.35,0.2)$.

\begin{table*}
\caption{Summary information about the archive images of NGC\,6791 used in this work.}
    \centering
    \begingroup
    \setlength{\tabcolsep}{12pt} 
    \renewcommand{\arraystretch}{1.2} 
    \begin{tabular}{cccccc}
    \hline \hline \\[-.3cm]
         DATE & N $\times$ EXPTIME & INSTRUMENT & FILTER & PROGRAM & PI   \\[.15cm]
         \hline \\[-.3cm] 
         2013 August 17 & $2\times700$\,s & UVIS/WFC3 & F275W & 13297 & G. Piotto \\
         2014 April 26 & $2\times707$\,s & UVIS/WFC3 & F275W & 13297 & G. Piotto \\
         2009 October 04 & $30\rm \,s+2\times400$\,s & UVIS/WFC3 & F336W & 13297 & G. Piotto \\
         2013 August 17 & $2\times300$\,s & UVIS/WFC3 & F336W & 13297 & G. Piotto \\
         2014 April 26 & $2\times297$\,s & UVIS/WFC3 & F336W & 13297 & G. Piotto \\
         2013 August 17 & 72\,s & UVIS/WFC3 & F438W & 13297 & G. Piotto \\
         2014 April 26 & 65\,s & UVIS/WFC3 & F438W & 13297 & G. Piotto \\
         2009 October 08 & 30\,s & UVIS/WFC3 & F814W & 11664 & G. Brown \\
         2009 October 07 & $2\times390$\,s & UVIS/WFC3 & F814W & 11664 & T. Brown \\
         2004 September 27 & $50\rm\,s+5\rm\,s$ & WFC/ACS & F814W & 10265 & T. Brown \\[.1cm]
         \hline \hline
    \end{tabular}
    \endgroup
    \label{dataset_table}
\end{table*}

\subsection{Simple-population clusters}
\label{subsec:sspcls}
To investigate the extended 1G sequence in simple-population clusters, we selected two star clusters, namely NGC\,6791 and NGC\,1783, with available multi-band HST photometry.

\begin{figure*}
     \centering
     \includegraphics[width=.9\textwidth]{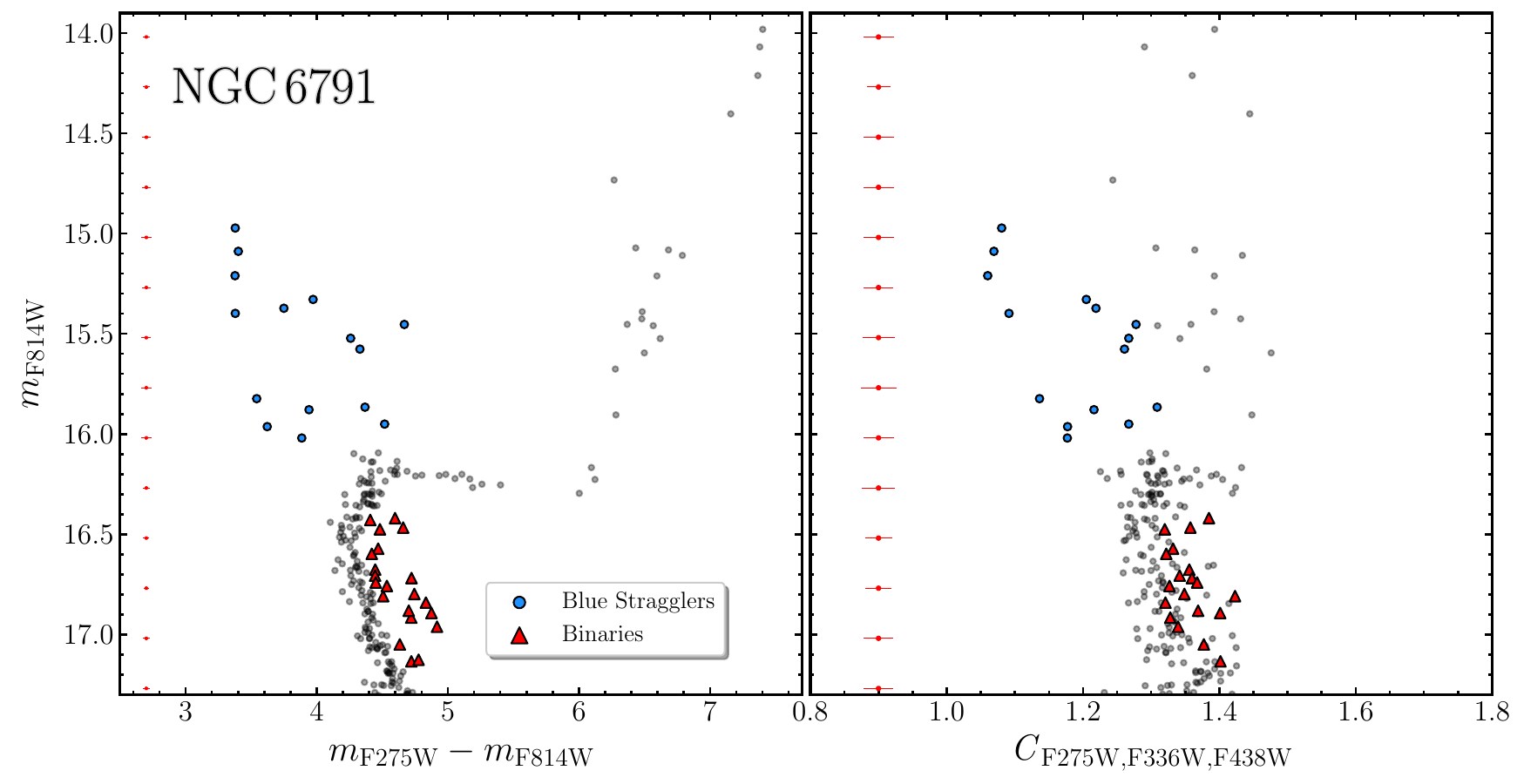}
     \caption{$m_{\rm F814W}$ vs.\,$m_{\rm F275W}-m_{\rm F814W}$ (left) CMD and $m_{\rm F814W}$ vs.\,$C_{\rm F275W,F336W,F438W}$ (right) pseudo-CMD of NGC\,6791. In both panels red triangles and blue points correspond to candidate binaries and blue stragglers, respectively, whereas the error bars represent photometric uncertainties.}
     \label{fig:ngc6791}
\end{figure*}

\begin{itemize}
\item NGC\,6791 is an old \citep[$\sim 8.5$\,Gyr,][]{brogard2012} and massive \citep[$\sim10^{4} M_{\odot}$, e.g.,][]{cordoni2023a} Galactic open cluster with a super-solar metallicity \citep[{[Fe/H]}$ \sim 0.3-0.4$, e.g.,][]{boesgaard2009}. In this work, we analyzed the innermost $\sim$ 2.7 $\times$ 2.7 arcmin$^{2}$ area of NGC\,6791 by using HST images collected with the F275W, F336W, F438W, and F814W filters of the UVIS/WFC3 and the F814W band of WFC/ACS. The main properties of these observations are summarized in Table~\ref{dataset_table}. Additionally, to extend our analysis outside the HST field of view, we combined the catalogs of proper motions and stellar positions provided by Gaia Data Release 3 \citep[][DR3]{gaiacoll2021} with the ground-based photometry from \citet{stetson2019a}. 

To derive the astro-photometric catalog for the central region of NGC\,6791 we used the computer program KS2 \citep[see, e.g.,][for more details]{sabbi2016, bellini2017, milone2023b}, which was developed by Jay Anderson as an evolution of the \texttt{kitchen\_sync} software \citep{anderson2008a}. Together with photometry, KS2 also provides a collection of diagnostic parameters that we exploited to select a sample of well-measured stars, as in \citet[][see their Sect.~2.4 for details]{milone2023b}. Finally, we corrected magnitudes for the effects of differential reddening following the method by \cite{milone2012a} and \cite{legnardi2023}.

The left panel of Fig.~\ref{fig:ngc6791} shows the resulting $m_{\rm F814W}$ vs.\,$m_{\rm F275W}-m_{\rm F814W}$ CMD of NGC\,6791 that we used to identify candidate binaries and blue stragglers. Similarly to photometric uncertainties, these objects contribute to the color broadening of evolutionary sequences in all CMDs. As an example, in the right panel of Fig.~\ref{fig:ngc6791} we plotted the $m_{\rm F814W}$ vs.\,$C_{\rm F275W,F336W,F438W}=(m_{\rm F275W}-m_{\rm F336W})-(m_{\rm F336W}-m_{\rm F438W})$ pseudo-CMD where the candidate binaries (red triangles) populate the reddest part of the MS, whereas blue stragglers (blue points) exhibit bluer colors than RGB stars. The $C_{\rm F275W,F336W,F438W}$ pseudo-color is sensitive to C, N, and O, thereby increasing the separation between stellar populations with different light-element abundances. The $C_{\rm F275W,F336W,F438W}$ color broadening is comparable to that of observational errors alone, indicated by the error bars on the right panel. This result allowed us to conclude that NGC\,6791 is a prototype of a simple-population cluster, thus corroborating previous results based on spectroscopy \citep[e.g.,][]{bragaglia2014, villanova2018} and on near-infrared photometry of M dwarfs \citep{dondoglio2022}.

To further investigate NGC\,6791 over a wide field of view we combined the photometry in the U, B, V, and I bands derived by Peter Stetson \citep{stetson2019a} together with the coordinates, proper motions, and parallaxes from the Gaia DR3 \citep{gaiacoll2021}. Photometry was obtained through an analysis conducted by Peter Stetson using images gathered from diverse ground-based telescopes. The procedures and computer programs outlined by \citet{stetson2005a} were employed, and the derived photometric data were calibrated against the reference system established by \citet{landolt1992}. We selected a sample of probable cluster members by using the proper motions and the stellar parallaxes from Gaia DR3 and the procedure by \citet{cordoni2018a}. Photometry has been corrected for differential reddening \citep[see][for details]{legnardi2023}.

\item NGC\,1783 is an intermediate-age \citep[$\sim 1.5$\,Gyr,][]{milone2023b} LMC cluster which shows no evidence of multiple stellar populations \citep{milone2020a}. We chose to analyze this object because it is the only simple-population MC cluster for which photometry in the UVIS/WFC3 filter F275W is available in the HST archive. In this work, we employed the catalog of NGC\,1783 by \cite{milone2023b}, which includes photometry in the F275W, F336W, F343N, and F438W bands of the UVIS/WFC3 and the F435W, F555W, and F814W filters of the ACS/WFC. Further details about the dataset and the procedures applied to reduce it can be found in the paper by Milone and collaborators.
\end{itemize}

To determine the photometric uncertainties in the HST dataset of NGC\,6791 and NGC\,1783 we carried out artificial-star (AS) tests following the method by \cite{anderson2008a}. In a nutshell, we generated a list including coordinates and fluxes for 100,000 stars. To do that, we assumed the same radial distribution for the ASs as the one measured for real stars. Moreover, we estimated the magnitudes of ASs by employing a set of fiducial lines derived from the observed CMDs.

To reduce ASs, we used once again KS2 and repeated the procedure described above. Subsequently, similar to our approach with real stars, we took advantage of the various diagnostic parameters provided by KS2 to select a sample of well-measured sources.

\subsection{Galactic globular clusters}
\label{subsec:galgcs}
\cite{legnardi2022} extensively investigated 1G stars in 55 Galactic GCs based on the ChMs of their RGB stars and inferred their internal metallicity variations. In this work, we further used the same dataset to derive the metallicity distributions of 1G stars and examine whether the color distribution of the 1G sequence is connected to some parameters of the host GCs. 

\section{An in-depth analysis of first-generation stars in 47 Tucanae}
\label{sec:47Tuc}
\begin{figure*}
    \centering
    \includegraphics[width=1.5\columnwidth]{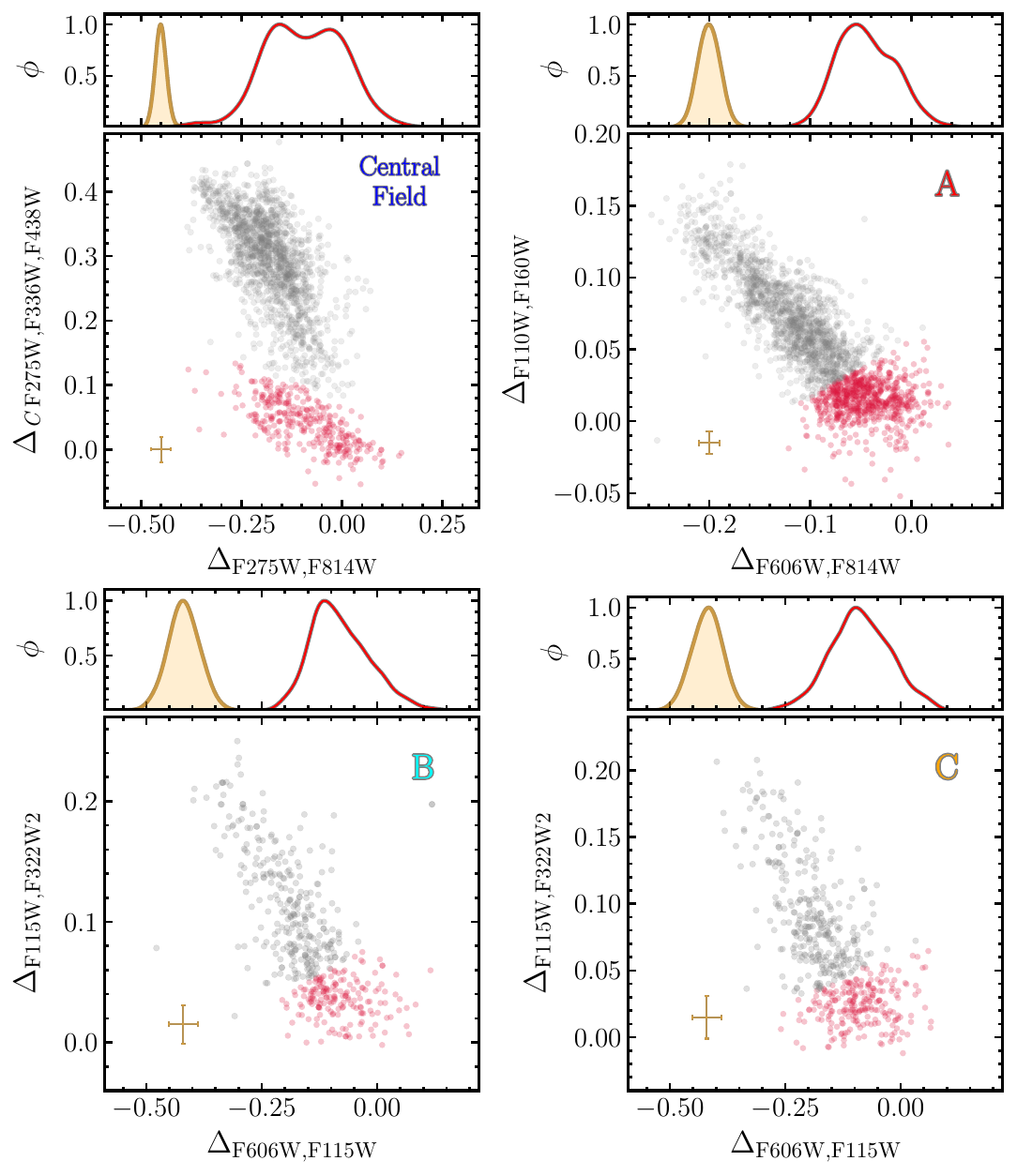}
    \caption{ChMs of 47\,Tucanae for the cluster center and the three external fields, namely A, B, and C. The ChM of the central region is based on RGB stars, whereas the ChMs of fields A, B, and C are derived for M dwarfs in the lower part of the MS. In all ChMs, 1G and 2G stars are marked with red and gray points, respectively, whereas photometric uncertainties are represented by the error bars in the bottom-left corner. The top panels illustrate the kernel-density distributions of 1G stars and observational errors.}
    \label{fig:47TucChMs}
\end{figure*}

To study 1G stars at progressively larger distances from the center of 47\,Tucanae we used the ChMs illustrated in Fig.~\ref{fig:47TucChMs}. The $\Delta_{\rm {\it C}\,F275W, F336W, F438W}$ vs.\,$\Delta_{\rm F275W,F814W}$ ChM of the central field is constructed for evolved RGB stars, whereas the $\Delta_{\rm F110W,F160W}$ vs.\,$\Delta_{\rm F606W,F814W}$ and the $\Delta_{\rm F115W,F322W2}$ vs.\,$\Delta_{\rm F606W,F115W}$ ChMs are derived for M dwarfs below the MS knee. 

As discussed in previous works \citep{milone2017, milone2023a}, the stars of all ChMs are distributed in two main populations, 1G and 2G, marked with red and gray colors, respectively. To identify 1G and 2G stars in the ChMs of RGB and M-dwarf stars, we adopted a similar selection as \citet[][see their Fig.~3]{milone2017} and \citet[][see panel b of their Figs.\,14 and B2]{milone2023a}, respectively\footnote{The $\Delta_{\rm F115W,F322W2}$ vs.\,$\Delta_{\rm F606W,F115W}$ ChM of field C closely resembles that of field B, hence we used the same criterion for both regions.}. In all ChMs, 1G stars lay in proximity of the ChM origin, whereas the 2G sequence extends towards high $\Delta_{\rm {\it C}\,F275W, F336W, F438W}$ ($\Delta_{\rm F110W,F160W}$, $\Delta_{\rm F115W,F322W2}$) and low $\Delta_{\rm F275W,F814W}$ ($\Delta_{\rm F606W,F814W}$, $\Delta_{\rm F606W,F115W}$) values, and exhibits hints of stellar overdensities, especially in the ChMs for M dwarfs. 

A visual inspection at Fig.~\ref{fig:47TucChMs} reveals that the 1G stars of 47\,Tucanae define an extended sequence along the x-axis of all ChMs. The width of the $\Delta_{\rm F275W,F814W}$ ($\Delta_{\rm F606W,F814W}$, $\Delta_{\rm F606W,F115W}$) pseudo-color for 1G stars is much wider than observational errors alone and is associated with star-to-star metallicity variations \citep{legnardi2022, marino2023}. Additionally, the pseudo-colors of 1G stars exhibit complex distributions, as suggested by the corresponding kernel-density distributions plotted in the upper panels of Fig.~\ref{fig:47TucChMs}. Clearly, the 1G kernel-density distribution of the central field shows two peaks located at $\Delta_{\rm F275W,F814W} \sim -0.18$ and $\sim -0.03$ mag, respectively. In the M-dwarf ChMs, instead, the frequency of 1G stars reaches its maximum at $\Delta_{\rm F606W,F814W} \sim -0.06$ and $\Delta_{\rm F606W,F115W} \sim -0.10$ mag, gradually declining towards the origin. The $\Delta_{\rm F110W,F160W}$ vs.\,$\Delta_{\rm F606W,F814W}$ ChM exhibits hints of an additional peak at $\Delta_{\rm F606W,F814W} \sim -0.01$ mag.
   
In the following, we further investigate the 1G stars of 47\,Tucanae. Specifically, in Sect.~\ref{subsec: 47Tuc_fe} we use the pseudo-color broadening observed along the x-axis of the ChMs to derive the metallicity distribution of 1G stars at distinct radial distances from the cluster center and investigate the radial behavior of metallicity variations within the 1G. In Sect.~\ref{subsec: 47Tuc_kin} we exploit the proper motion catalogs introduced in Sect.~\ref{sec:data} to study the internal kinematics of 1G stars.

\subsection{The radial distribution of metallicity variations among first-generation stars}
\label{subsec: 47Tuc_fe}
To infer the [Fe/H] distribution of 1G stars at different radial distances from the core of 47\,Tucanae we applied a similar approach to that described in \cite{legnardi2022}, which is based on the method originally developed by \citeauthor{cignoni2010} (\citeyear{cignoni2010}; see also \citealt{cordoni2022}) to derive the age of young star clusters through the MS turn-on.

First, we generated a grid of simulated ChMs, where all stars share the same helium and light-element contents but have different metallicities. In close analogy with \cite{legnardi2022}, we simulated N stellar populations with [Fe/H] ranging from [Fe/H]$_{0} -0.5$ to [Fe/H]$_{0} +0.5$, where [Fe/H]$_{0}=-0.72$ \citep[from][2010 update]{harris1996}. Then, we combined the simulated ChMs to derive the $\Delta_{\rm F275W,F814W}$ ($\Delta_{\rm F606W,F814W}$, $\Delta_{\rm F606W,F115W}$) histogram distribution. To do that, we multiplied the ChM of each population $j$ for $c_{j}$, a scaling coefficient that ranges from 0 to 1 and tells how much the population $j$ contributes to the final combined $\Delta_{\rm F275W,F814W}$ ($\Delta_{\rm F606W,F814W}$, $\Delta_{\rm F606W,F115W}$) distribution. Finally, we compared the resulting simulated histogram with the observed one by minimizing the Poissonian $\chi^{2}$ given by the equation:
\begin{equation}
     \chi^{2}=\sum_{\rm i}^{N_{\rm bins}} n_{\rm i} ln (n_{\rm i}/m_{\rm i}) - n_{\rm i} + m_{\rm i}
\end{equation}
where $n_{\rm i}$ and $m_{\rm i}$ are the bin values of the observed and simulated 1G stars, respectively. For the $\chi^{2}$ minimization, as in \cite{cordoni2022} and \cite{legnardi2022}, we used the \texttt{geneticalgorithm} Python public library\footnote{\url{https://pypi.org/project/geneticalgorithm/}}. 

\begin{figure*}
    \centering
    \includegraphics[width=1.5\columnwidth]{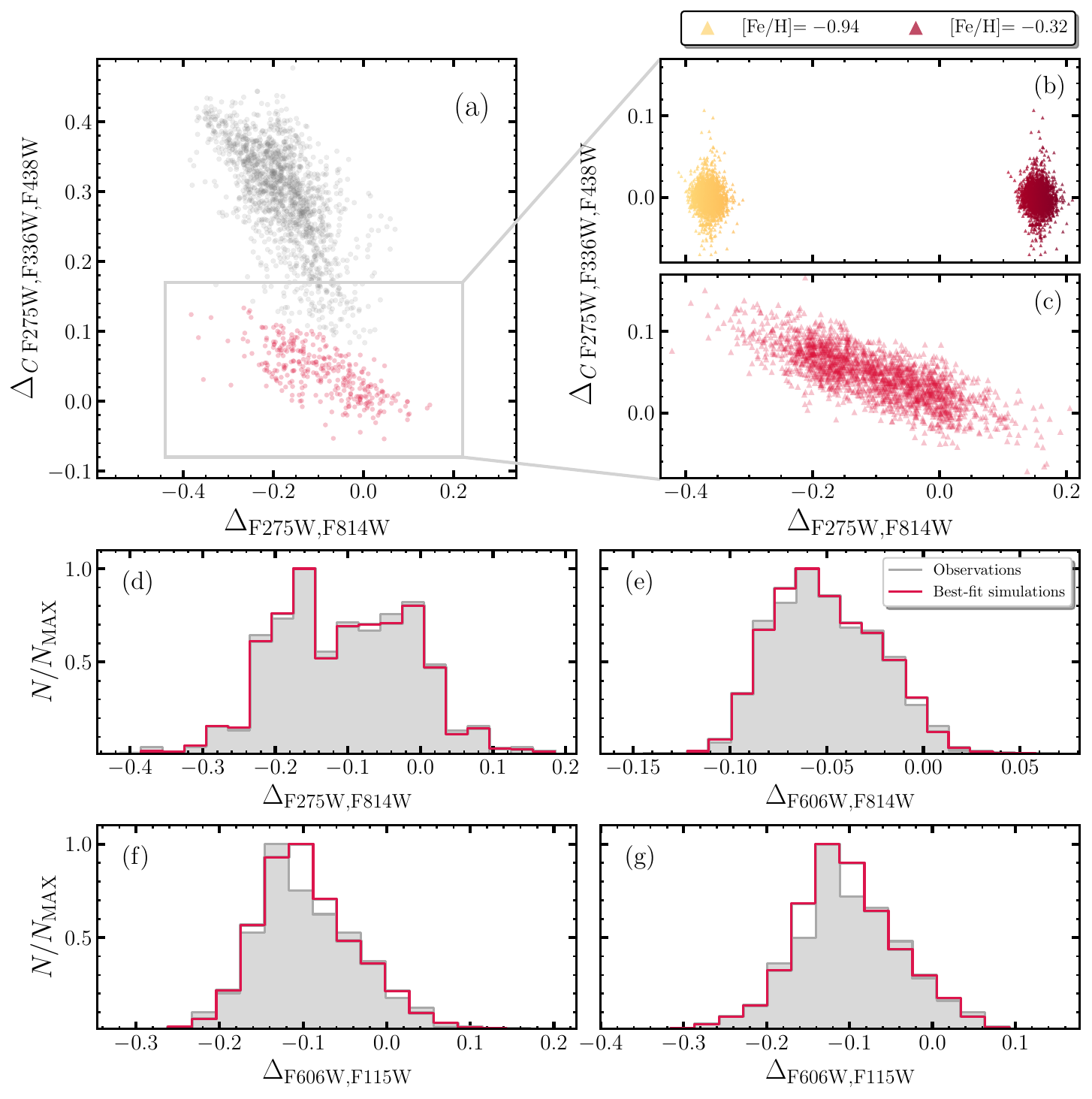} 
    \caption{{\it Top panels.} Comparison between the observed $\Delta_{\rm {\it C}\,F275W, F336W, F438W}$ vs.\,$\Delta_{\rm F275W,F814W}$ ChM for 1G stars within the central field (red points in panel a) and the ChM of simulated 1G stars (panels b and c). All stars in the ChMs of panel b have the same iron abundances of [Fe/H]$=-0.94$ and $-0.32$, respectively, whereas the ChM of panel d represents the best-fitting simulation. The number of simulated 1G stars is $\sim 5$ times larger than the observed ones. {\it Bottom panels.} Histograms of the observed $\Delta_{\rm F275W,F814W}$ ($\Delta_{\rm F606W,F814W}$, $\Delta_{\rm F606W,F115W}$) distributions inferred from 1G stars within the cluster center (panel d), fields A, B, and C (panels e, f, and g, respectively). In each panel, the red line corresponds to the simulated 1G $\Delta_{\rm F275W,F814W}$ ($\Delta_{\rm F606W,F814W}$, $\Delta_{\rm F606W,F115W}$) distribution that provides the best match with the observed data.}
    \label{fig:histsim}
\end{figure*}

As an example, the top panels of Fig.~\ref{fig:histsim} compare the ChM of 1G stars in the central field (red points in panel a) with the ChMs of simulated 1G stars (panels b and c).  All stars in the ChMs of panel b share the same iron abundances of [Fe/H]$=-0.94$ and $-0.32$, respectively, whereas the ChM plotted in panel c provides the best match with the observed one. The comparison between the observed $\Delta_{\rm F275W, F814W}$ and the best-fitting $\Delta_{\rm F275W, F814W}$ pseudo-color distribution is illustrated in panel d. Similar comparisons for 1G stars within fields A, B, and C are provided in panels e, f, and g, respectively.

\begin{figure}
    \centering
    \includegraphics[width=.95\columnwidth]{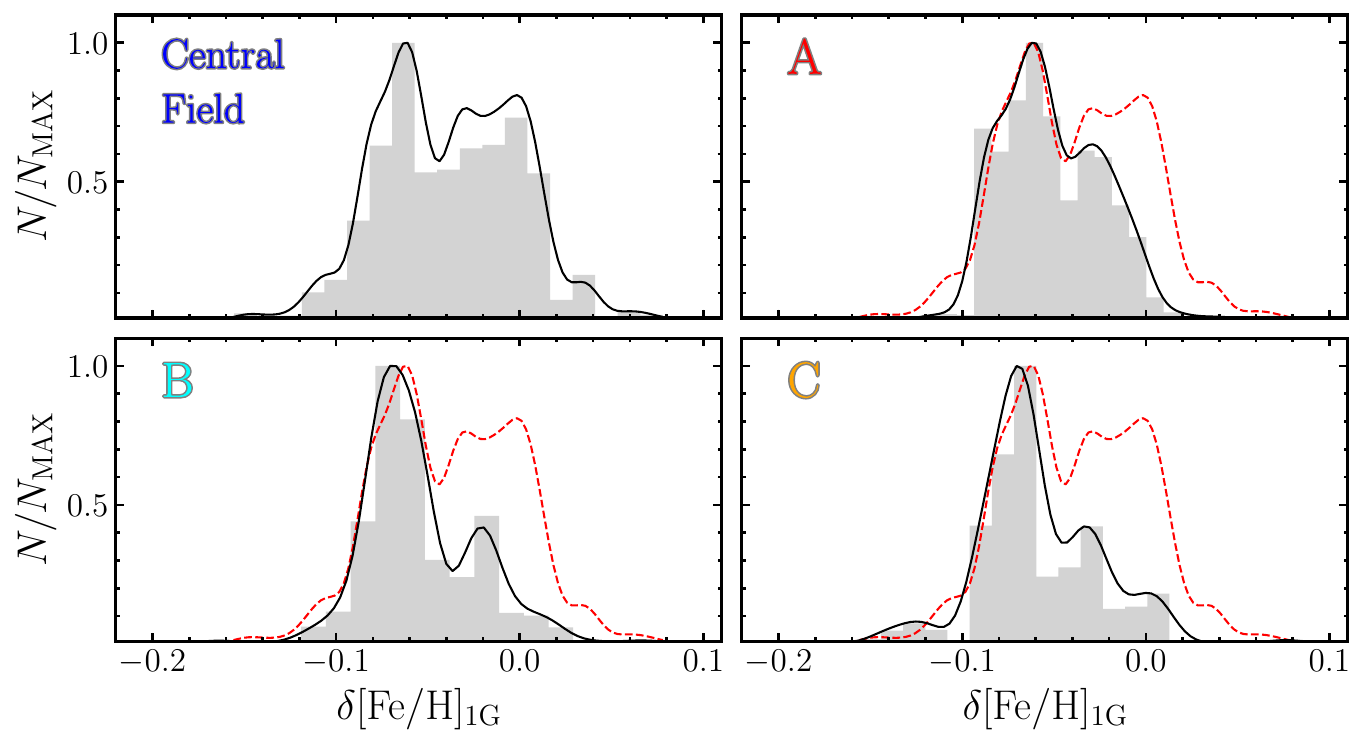}
    \caption{Histograms of the 1G metallicity distribution inferred from RGB stars within the cluster center (top-left) and M dwarfs of fields A (top-right), B (bottom-left), and C (bottom-right). In each panel, the corresponding kernel-density distribution is superimposed to the histograms with the continuous black line, whereas the red dashed line represents the kernel-density distribution derived for RGB stars in the cluster core.}
    \label{fig:irondist}
\end{figure}

Figure~\ref{fig:irondist} illustrates the resulting histograms of [Fe/H] variations among 1G stars in the cluster center, fields A, B, and C, together with their corresponding kernel-density distributions. The [Fe/H] distributions for red giants and M dwarfs span similar intervals of $\sim 0.1$ dex, ranging from $\delta$[Fe/H]$_{\rm 1G} \sim -0.15$ to $\sim 0.05$, and exhibit a main peak at $\delta$[Fe/H]$_{\rm 1G} \sim -0.07$. 

To compare the cluster center with the external regions, in panels A, B, and C we plotted the kernel-density distribution of the relative iron abundances of 1G stars within the central field (red-dashed line). Clearly, the latter shows a secondary peak centered at $\delta$[Fe/H]$_{\rm 1G} \sim -0.02$ that gradually becomes less evident moving towards the outer fields. 

\begin{figure}
    \centering
    \includegraphics[width=.95\columnwidth]{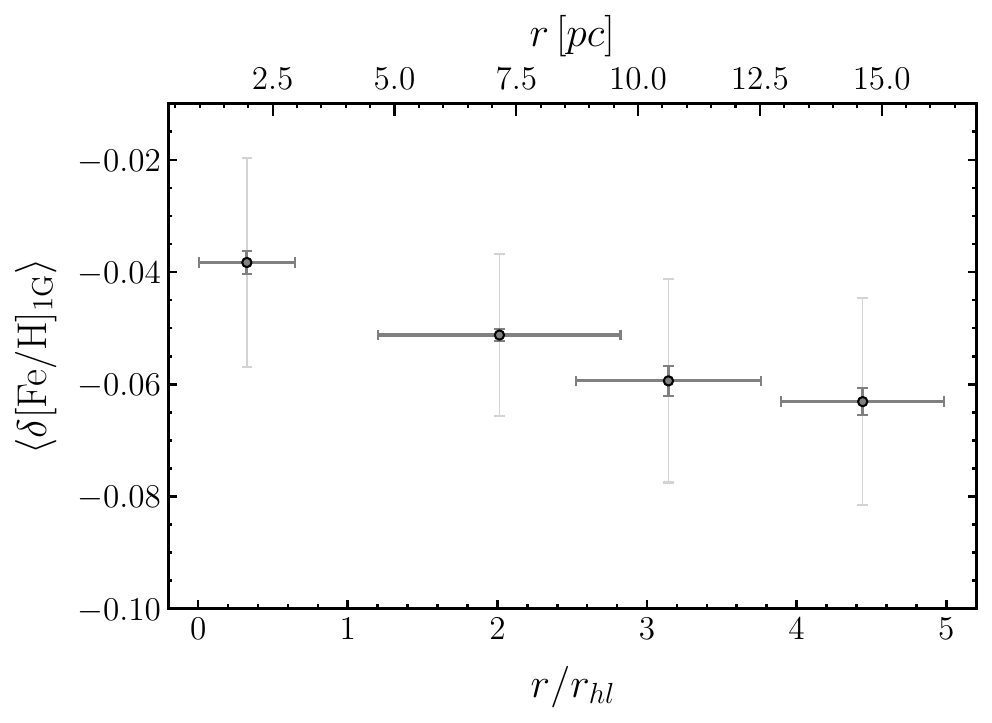}
    \caption{Average metallicity variations among 1G stars, $\langle \delta$[Fe/H]$_{\rm 1G} \rangle$, as a function of the radial distance from the cluster center. Gray points correspond to the average $\delta$[Fe/H] of 1G stars within each analyzed field, whereas the thick and thin error bars indicate the uncertainties associated with $\langle \delta$[Fe/H]$_{\rm 1G} \rangle$ and the dispersions, respectively. On the bottom axis, the distance from the cluster center is normalized to the half-light radius from \citet[][$r_{hl}=2.78$ arcmin]{baumgardt2018}. On the top axis, instead, the radial coordinate is converted to parsec by adopting a distance of 4.41 kpc \citep{baumgardt2018}. The horizontal bars mark the extension of each radial interval.}
    \label{fig:ironraddist}
\end{figure}

To further investigate the 1G metallicity variations within the cluster, we calculated the average value of the $\delta$[Fe/H]$_{\rm 1G}$ distribution, $\langle \delta$[Fe/H]$_{\rm 1G} \rangle$, for each of the four analyzed fields, that we plotted as a function of the radial coordinate in Fig.~\ref{fig:ironraddist}. On the bottom axis, we normalized the radial coordinate to the half-light radius from \citet[][$r_{hl}=2.78$]{baumgardt2018}. Conversely, on the top axis, we converted the radial coordinate in parsec by assuming a distance of 4.41 kpc \citep{baumgardt2018}. 

The radial distribution of $\langle \delta$[Fe/H]$_{\rm 1G} \rangle$ suggests a concentration of more metal-rich 1G stars in the innermost cluster regions, whereas in the outer fields, 1G stars have lower average iron content. To demonstrate that this result is not affected by the adopted 1G samples, we repeated the analysis based on 1G stars identified in ChMs that maximize the separation between distinct stellar populations. Our analysis confirmed that we obtained consistent results when deriving [Fe/H] variations from these new samples of 1G stars. As an example, for M dwarfs within field A, we selected 1G stars in the $\Delta_{\rm {\it C}\,F606W,F814W,F322W2}$ vs.\,$\Delta_{\rm F606W,F814W}$ ChM \citep[see Fig.~8 of][]{marino2024a}, finding that the average [Fe/H] variation differs just by $\sim 0.001$ dex from the original value. Similar results have been found for field C, where we used the $\Delta_{\rm  F110W,F160W,F115W,F322W2}$ vs.\,$\Delta_{\rm F606W,F115W}$ ChM \citep[see Fig.~9 of][]{marino2024a} to identify a new sample of 1G stars. 

However, it is important to note that the outcomes for the central field are derived from RGB stars, while in the other fields, we focused on analyzing M dwarfs. Consequently, it cannot be ruled out that the observed difference in metallicity between the central and external fields may be influenced by the varying stellar masses within these regions.

\subsection{Internal kinematics of first-generation stars}
\label{subsec: 47Tuc_kin}
\begin{figure}
    \centering
    \includegraphics[width=.95\columnwidth]{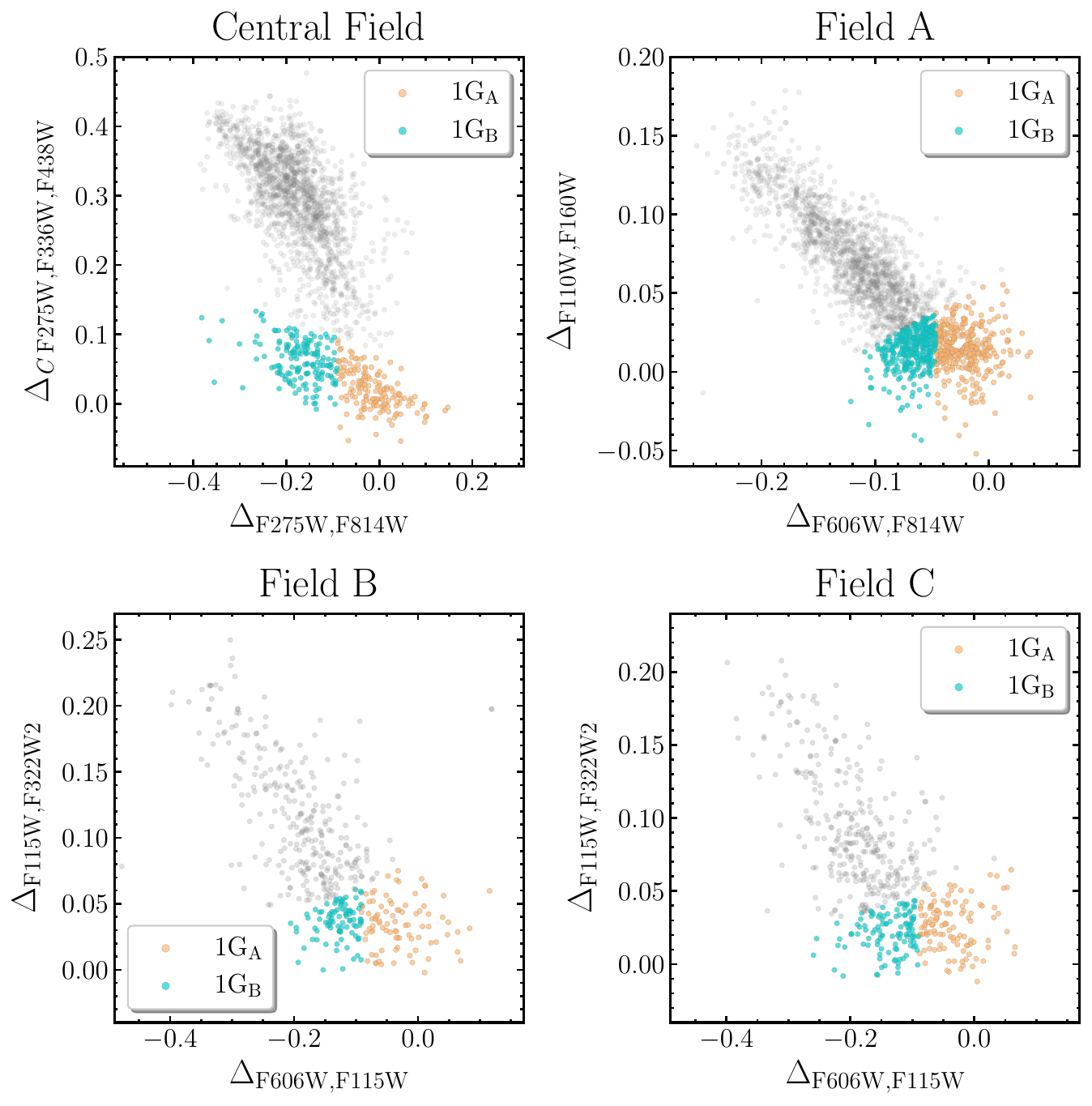}
    \caption{Reproduction of the ChMs of Fig.~\ref{fig:47TucChMs} where we have highlighted with orange and cyan the sub-samples of 1G$_{\rm A}$ and 1G$_{\rm B}$ stars, respectively.}
    \label{fig:47TucChMs_1g}
\end{figure}

To investigate the 1G kinematics, we first identified two groups of 1G stars based on the direction of their color extension in the ChM, with the criteria that each sample includes approximately half of the 1G members. The two sub-groups of 1G stars, namely 1G$_{\rm A}$ and 1G$_{\rm B}$, are marked with orange and cyan points in the ChMs of Fig.~\ref{fig:47TucChMs_1g}. We then computed the radial ($\sigma_{\rm RAD}$) and tangential ($\sigma_{\rm TAN}$) velocity dispersions of 1G$_{\rm A}$ and 1G$_{\rm B}$ following the procedure described in \citeauthor{cordoni2023b}(\citeyear{cordoni2023b}; see also \citealp{cordoni2020b,cordoni2020a}), which is based on the minimization of the log-likelihood function by \cite{bianchini2018}. We estimated the corresponding uncertainties through the MCMC algorithm \texttt{emcee} \citep{fm2013}, as in \cite{cordoni2023b}.

\begin{figure}
    \centering
    \includegraphics[width=.95\columnwidth]{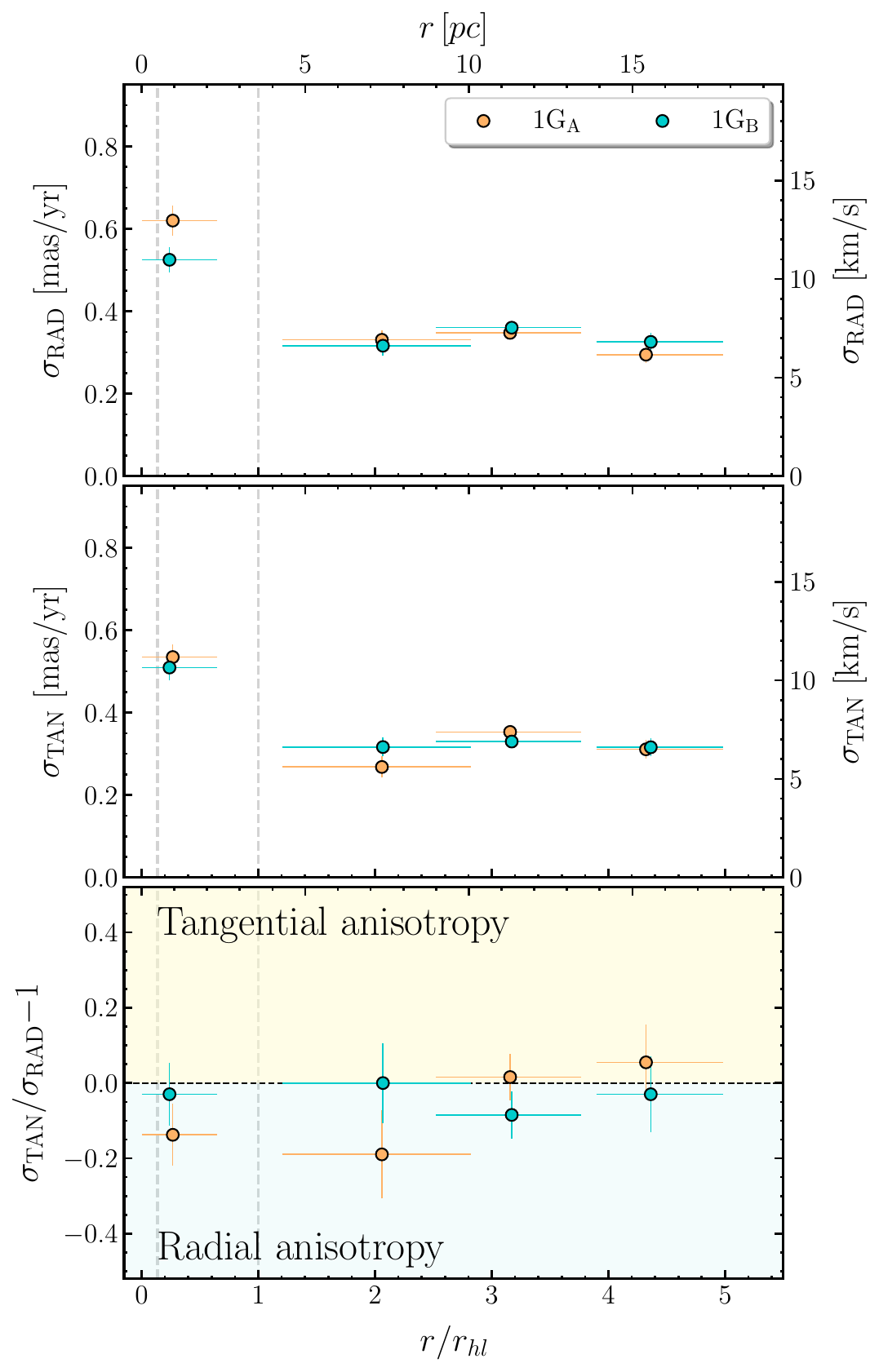}
    \caption{This figure illustrates the velocity dispersion profile along the radial (top panel) and tangential (middle panel) direction, together with the corresponding anisotropy profile (bottom panel). In all the panels, orange and cyan points indicate 1G$_{\rm A}$ and 1G$_{\rm B}$ stars, respectively, while the gray dashed lines correspond to the core and the half-light radius. On the bottom axis, the distance from the cluster center is normalized to the half-light radius from \citet[][$r_{hl}=2.78$ arcmin]{baumgardt2018}. On the top axis, instead, the radial coordinate is converted to parsec by adopting a distance of 4.41 kpc \citep{baumgardt2018}. The horizontal bars mark the extension of each radial interval.}
    \label{fig:kin}
\end{figure}

The resulting radial and tangential velocity dispersion profiles are illustrated in the upper and middle panels of Fig.~\ref{fig:kin}, respectively. Additionally, we calculated the anisotropy parameter, $\beta = \sigma_{\rm TAN}/\sigma_{\rm RAD}-1$, that we show as a function of the radial coordinate in the lower panel of Fig.~\ref{fig:kin}, where the horizontal dashed line corresponds to an isotropic stellar system. In each panel, orange and cyan points correspond to 1G$_{\rm A}$ and 1G$_{\rm B}$ stars, respectively, while on the bottom axis, the distance from the cluster center is normalized to the half-light radius from \citet[][$r_{hl}=2.78$ arcmin]{baumgardt2018}. On the top axis, instead, we converted the radial coordinate in parsec by assuming a distance of 4.41 kpc \citep{baumgardt2018}. To ensure having enough statistics, we treated each of the analyzed fields as one circular bin so that the resulting dynamical profiles are populated by four points for each 1G group. The horizontal bars associated with each point indicate the extension of the corresponding radial interval. 

Our analysis of 1G stars in 47\,Tucanae reveals that 1G$_{\rm A}$ and 1G$_{\rm B}$ stars share similar radial and tangential velocity dispersion profiles, ranging from $\sim 0.6$ mas/yr ($\sim 13$ km/s) close to the cluster center to $\sim 0.25$ mas/yr ($\sim 5$ km/s) at a radial distance of $\sim 4.5$ $r_{hl}$. The only exception is represented by the innermost region, where the radial velocity dispersion of 1G$_{\rm A}$ stars is slightly larger than that of 1G$_{\rm B}$ members. The anisotropy profile in the bottom panel of Fig.~\ref{fig:kin} demonstrates that 1G stars exhibit, on average, an isotropic motion, thus confirming previous results in the literature \citep{richer2013,milone2018a,cordoni2020a}. Moreover, 1G$_{\rm A}$ and 1G$_{\rm B}$ stars show no significant difference in their level of anisotropy.

To test whether our selection of 1G sub-samples affected the results, we identified two new groups of 1G$_{\rm A}$ and 1G$_{\rm B}$ stars according to the metallicity distribution derived in Sect.~\ref{subsec: 47Tuc_fe}. Then, we recomputed the velocity dispersion and anisotropy profiles finding consistent results with the original ones. Overall, metal-rich and metal-poor 1G stars exhibit isotropic motions with no significant differences in their radial and tangential velocity dispersion profiles.

The fact that the 1G groups of 47\,Tucanae have different radial distributions but similar kinematics would deserve further investigation. Indeed, stellar populations with different radial distributions typically exhibit different kinematics \citep[see, e.g.,][for the case of 1G and 2G stars in 47\,Tucanae]{richer2013, milone2018a,cordoni2020a}. More data are needed to better constrain the kinematics of 1G stars with different metallicities, whereas dynamic simulations are crucial to shed light on the initial configuration and the dynamic evolution of metal-rich and metal-poor 1G stars in 47\,Tucanae.

\begin{figure*}
    \centering
    \includegraphics[width=.85\textwidth]{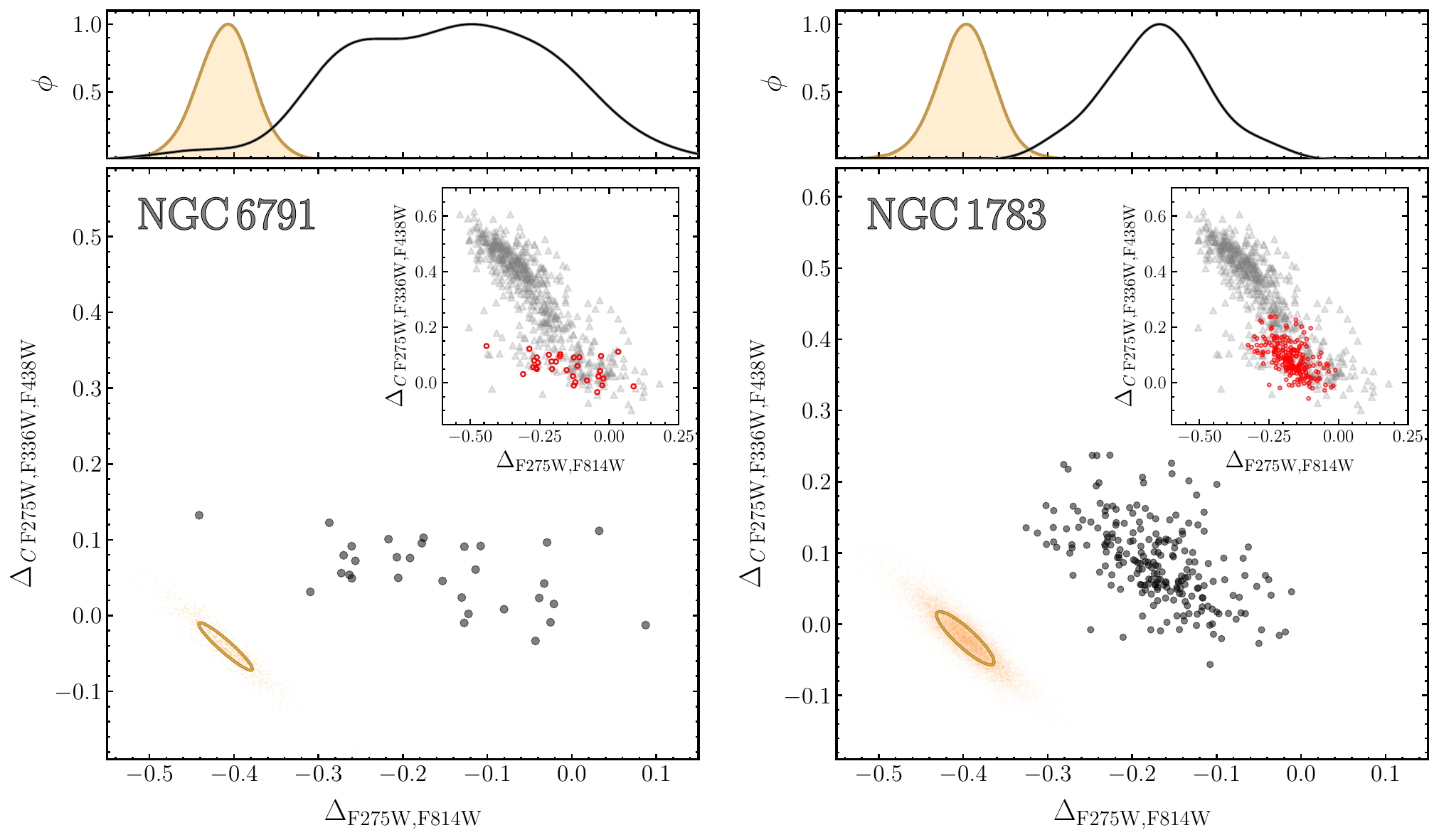}       
    \caption{$\Delta_{\rm {\it C}\,F275W, F336W, F438W}$ vs.\,$\Delta_{\rm F275W,F814W}$ ChMs for NGC\,6791 (left) and NGC\,1783 (right). In both panels, the orange points in the bottom-left corner indicate the distribution expected for a simple population derived from ASs, while the orange ellipses encircle the 68.27$\%$ of simulated points. The $\Delta_{\rm F275W,F814W}$ kernel-density distributions of observed and simulated stars are colored in black and orange, respectively, in the top panel of each figure. In the insets, we show a comparison between the ChMs of the two simple-population clusters and the one of the metal-rich GC NGC\,6624 \citep{milone2017}.}
    \label{fig:SSPs_ChMs}
\end{figure*}

\begin{figure*}
    \centering
    \includegraphics[width=.95\textwidth]{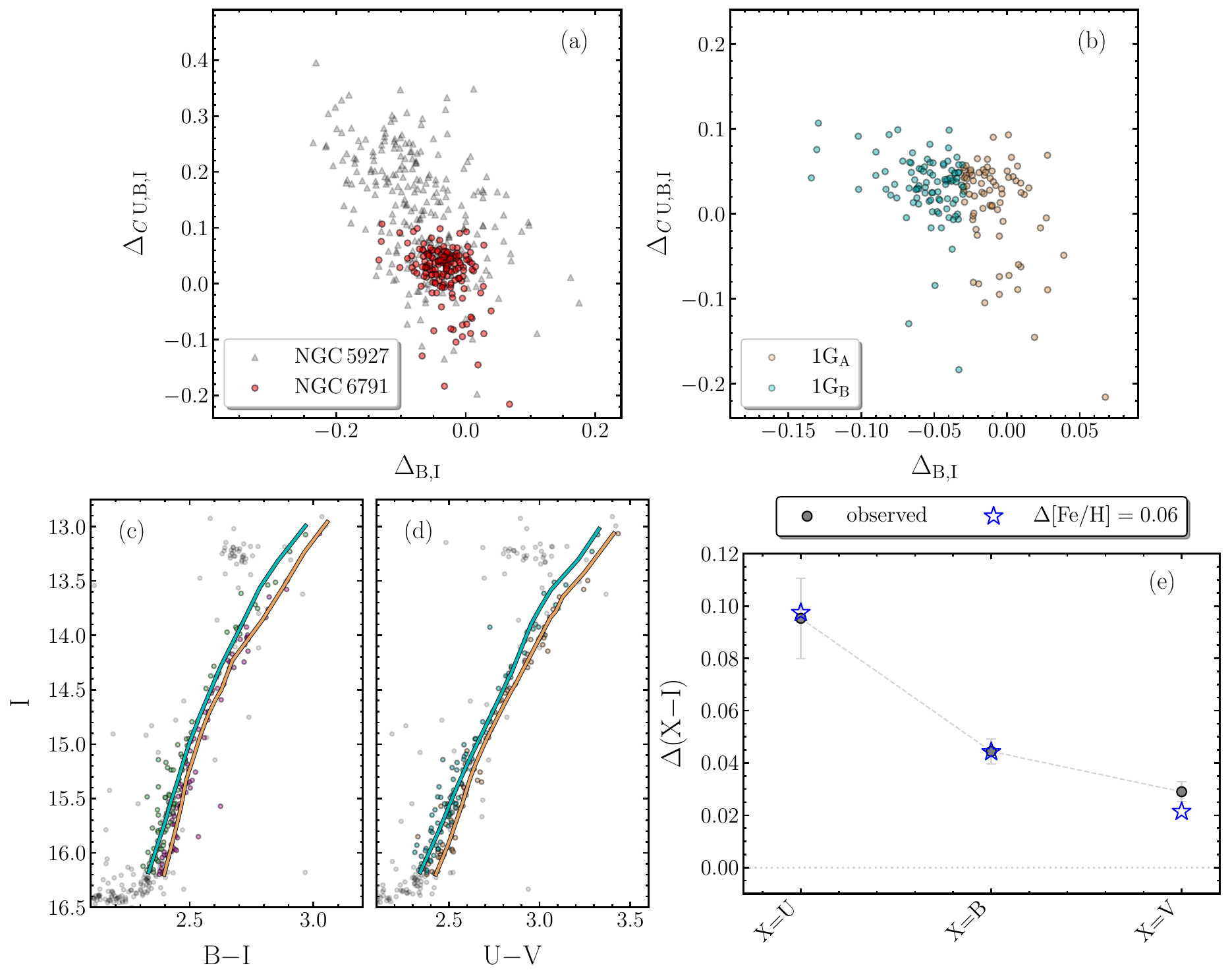}
    \caption{{\it Panel a.} $\Delta_{\rm {\it C}\,U,B,I}$ vs.\,$\Delta_{\rm B,I}$ ChMs of NGC\,6791 (red points) and the metal-rich GC NGC\,5927 \citep[gray triangles,][]{jang2022a}. {\it Panel b.} Reproduction of the ground-based ChM of NGC\,6791 where we marked with orange and cyan points the two sub-samples of 1G$_{\rm A}$ and 1G$_{\rm B}$ stars, respectively. {\it Panels c and d.} I vs.\,B$-$I and I vs.\,U$-$V CMDs of RGB stars in NGC\,6791, where the fiducial lines for 1G$_{\rm A}$ and 1G$_{\rm B}$ are plotted in orange and cyan, respectively. {\it Panel e.} $\Delta$(X$-$I) (where X=U, B, and V) calculated at I$^{ref}=15.0$, against the central wavelength of the X filter. The colors inferred from the best-fitting isochrones are overplotted with blue-starred symbols. See the text for details.}
    \label{fig:6791ground}
\end{figure*}

\section{Simple-population clusters: NGC 6791 and NGC 1783}
\label{sec:SSPCs}
So far, the color broadening of the 1G sequence has been detected only in Galactic GCs with multiple stellar populations. To verify whether this phenomenon depends on the occurrence of multiple stellar populations, in the following we analyze two simple-population clusters, namely NGC\,6791 and NGC\,1783. 

\subsection{The extended stellar population of NGC 6791 and NGC 1783}
\label{subsec: SSSPCs_ChMS}
To investigate whether chemical anomalies similar to the ones detected within the 1G of Galactic GCs are common also in simple-population clusters, we derived the $\Delta_{\rm {\it C}\,F275W, F336W, F438W}$ vs.\,$\Delta_{\rm F275W,F814W}$ ChMs of NGC\,1783 and NGC\,6791 by extending the method by \citet{milone2017} to the two analyzed targets.

As an example, to derive the ChM of NGC\,1783 we first selected RGB stars with $16.5\le m_{\rm F814W} \le18.5$ mag. Then, we defined the blue and red boundaries of the RGB in the $m_{\rm F814W}$ vs.\,$C_{\rm F275W,F336W,F438W}$ pseudo-CMD and the $m_{\rm F814W}$ vs.\,$m_{\rm F275W}-m_{\rm F814W}$ CMD. To do that, we divided the sample of selected stars in magnitude bins of size 0.2 mag and for each bin we calculated the 4$^{\rm th}$ and the 96$^{\rm th}$ percentile of the $m_{\rm F275W}-m_{\rm F814W}$ color and $C_{\rm F275W,F336W,F438W}$ pseudo-color distribution. These quantities have been interpolated with the mean $m_{\rm F814W}$ magnitude of each bin to derive the fiducial lines that mark the blue and red RGB envelopes. Finally, we took advantage of Eqs.~(1) and (2) of \citet{milone2017} to compute the $\Delta_{\rm F275W,F814W}$ and $\Delta_{\rm {\it C}\,F275W, F336W, F438W}$ pseudo-color. A similar procedure has been used to derive the ChM of NGC\,6791, but in this case we considered RGB stars with $13.4\le m_{\rm F814W} \le16.4$ mag.

Results are illustrated in the left and right panels of Fig.~\ref{fig:SSPs_ChMs}, where we show the ChMs of NGC\,6791 and NGC\,1783, respectively. In the bottom-left corner of each panel, we plot the distribution expected for a simple stellar population, derived by using ASs, and the ellipse that includes the $68.27 \%$ of the simulated points. Additionally, the top panels of each figure report the $\Delta_{\rm F275W,F814W}$ kernel-density distribution of observed (black line) and simulated (orange line) stars.

Similarly to the 1G sequences of Galactic GCs, which move from the origin of the ChM to high $\Delta_{\rm {\it C}\,F275W, F336W, F438W}$ and low $\Delta_{\rm F275W,F814W}$ values, the ChMs of NGC\,6791 and NGC\,1783 exhibit a single stellar population that starts in proximity of ($\Delta_{\rm F275W,F814W}, \Delta_{\rm {\it C}\,F275W, F336W, F438W})\sim(0.00,0.00)$ and extends to ($\Delta_{\rm F275W,F814W}, \Delta_{\rm {\it C}\,F275W, F336W, F438W})\sim(-0.40,0.15)$ and ($\Delta_{\rm F275W,F814W}, \Delta_{\rm {\it C}\,F275W, F336W, F438W})\sim(-0.30,0.25)$, respectively. The insets of Fig.~\ref{fig:SSPs_ChMs} reinforce this visual impression, as the ChMs of the two investigated targets overlap with the 1G sequence of NGC\,6624, which is one of the most metal-rich Galactic GCs \citep[{[Fe/H]}$=-0.44$ dex,][2010 updated]{harris1996} with available ChM.

As suggested by the corresponding kernel-density distribution in the top panels of Fig.~\ref{fig:SSPs_ChMs}, the ChMs of both clusters display an extended sequence in the direction of the $\Delta_{\rm F275W,F814W}$ pseudo-color. To estimate the contribution of observational errors to the observed color spread, we quantified the $\Delta_{\rm F275W,F814W}$ color spread of observed and simulated stars in the ChMs of the two investigated targets. We found that the observed color spread of NGC\,6791 ($\sigma^{\rm obs}_{\rm F275W,F814W}$=0.129$\pm$0.018 mag) and NGC\,1783 ($\sigma^{\rm obs}_{\rm F275W,F814W}$=0.058$\pm$0.003 mag) are larger than the corresponding color elongation of a simulated simple stellar population ($\sigma^{\rm sim}_{\rm F275W,F814W}$=0.030 mag and 0.032 mag, respectively). 

To increase the number of available stars in NGC\,6791 and study 1G RGB stars within 10 arcmin from the cluster center, we derived the $\Delta_{\rm {\it C}\,U,B,I}$ vs.\,$\Delta_{\rm B,I}$ ChM from ground-based photometry, which is based on the ${\rm B}-{\rm I}$ color and the $C_{\rm U,B,I}= {(\rm U}-{\rm B})-({\rm B}-{\rm I})$ pseudo-color \citep{jang2022a}. A visual inspection at panel a of Fig.~\ref{fig:6791ground}, where we show the resulting ChM superimposed to the one of NGC\,5927  \citep[{[Fe/H]}$=-0.49$ dex,][2010 updated]{harris1996} derived by \citet{jang2022a}, reveals that NGC\,6791 exhibits a single stellar population which aligns with the 1G of NGC\,5927. This result allowed us to confirm that NGC\,6791 is a simple-population cluster.

To further investigate the presence of metallicity variations in NGC\,6791, we extended to these clusters the procedure used by \citet[][see also \citealp{anderson2009}]{milone2010} to demonstrate that NGC\,6752 hosts multiple populations. As illustrated in panel b of Fig.~\ref{fig:6791ground}, we first identified two groups of stars in the $\Delta_{\rm {\it C}\,U,B,I}$ vs.\,$\Delta_{\rm B,I}$ ChM, namely 1G$_{\rm A}$ and 1G$_{\rm B}$, with the criteria that each sample includes about half of the total number of stars. We then analyzed the distribution of these two groups of stars in CMDs constructed with combinations of filters that are not used to derive the ChM. Panels c and d of Fig.~\ref{fig:6791ground} show the I vs.\,B$-$I and the I vs.\,U$-$V CMDs where we used orange and cyan colors to identify 1G$_{\rm A}$ and 1G$_{\rm B}$ stars, respectively. As highlighted by the corresponding fiducial lines, the two groups of stars remain well separated in both diagrams, thus corroborating the conclusion that the color differences between 1G$_{\rm A}$ and 1G$_{\rm B}$ stars with similar luminosity are intrinsic. 

To quantify the metallicity difference between the two selected groups of NGC\,6791 stars, we compared the observed stellar colors with the solar-scaled isochrones from \citet{dotter2008} that provide the best fit to the I vs.\,B$-$I CMD. Specifically, we assumed metallicity corresponding to [Fe/H]=0.4 dex, age=8.5 Gyr, distance modulus, (m$-$M)$_{0}$=13.5 mag, and reddening, $E(B-V)$=0.1 mag. We selected the two groups of 1G stars identified in panel b of Fig.~\ref{fig:6791ground} in the I vs.\,X$-$I CMDs, where X=U, B, and V, and we derived the corresponding fiducial lines. Subsequently, we calculated the $\Delta$(X$-$I) color difference between the 1G$_{\rm A}$ and 1G$_{\rm B}$ fiducials at a reference magnitude of I$^{ref} =$15.0 mag. Results are illustrated in panel e of Fig.~\ref{fig:6791ground}, where we compare the observed color differences between 1G$_{\rm A}$ and 1G$_{\rm B}$ stars with those derived from the best-fitting isochrones. We found that the selected groups of NGC\,6791 stars are consistent with a difference in iron abundance of $\Delta$[Fe/H]$ \sim 0.06$ dex, which provides a lower limit to the maximum iron variations within NGC\,6791. In summary, results from ground-based photometry of NGC\,6791 confirm the conclusions that we obtained from HST data that this cluster is composed of 1G stars alone, which are not chemically homogeneous. 
   
Our analysis on NGC\,6791 and NGC\,1783 suggests that the extension of the sequence observed on the ChM of the two simple-population clusters has an intrinsic origin, which is related to star-to-star metallicity variations. Therefore, this phenomenon is not necessarily associated with the occurrence of multiple stellar populations in GCs.

\subsection{Metallicity variations within NGC 6791 and NGC 1783}
\label{subsec:metSSPs}
Recent work, based on the $m_{\rm F275W}-m_{\rm F814W}$ color width of RGB stars (W$^{\rm 1G}_{\rm F275W,F814W}$) of 55 Galactic GCs, reveals that the maximum [Fe/H] variations among 1G stars, $\delta$[Fe/H]$_{\rm 1G}^{\rm MAX}$, range from less than $\sim 0.05$ to $\sim 0.30$ dex \citep{legnardi2022}. In the following, we estimate the maximum [Fe/H] spreads of NGC\,6791 and NGC\,1783 by adapting the procedure of \citet{legnardi2022} to the two investigated targets. 

For each cluster, we first computed the intrinsic $\Delta_{\rm F275W,F814W}$ pseudo-color extension. To do that, in close analogy with \citet{milone2017}, we subtracted the color errors in quadrature to the observed $\Delta_{\rm F275W,F814W}$ pseudo-color width, defined as the difference between the 90$^{\rm th}$ and the 10$^{\rm th}$ percentile of the $\Delta_{\rm F275W,F814W}$ distribution. As suggested by the visual comparison of the ChMs in Fig.~\ref{fig:SSPs_ChMs}, the intrinsic width of NGC\,6791 (W$^{\rm 1G}_{\rm F275W,F814W}=0.266\pm0.032$ mag) is consistently larger than that of NGC\,1783 (W$^{\rm 1G}_{\rm F275W,F814W}=0.149\pm0.014$ mag). 

Subsequently, we used the relation between the $m_{\rm F275W}-m_{\rm F814W}$ color and metallicity from \citet{dotter2008} to transform the $\Delta_{\rm F275W,F814W}$ color extension in [Fe/H] variations, in the assumption that the color spread observed on the ChM is entirely caused by iron spreads. We found that NGC\,6791 exhibits moderate [Fe/H] variations ($\delta$[Fe/H]$_{\rm 1G}^{\rm MAX}=0.080\pm0.011$), whereas NGC\,1783 is characterized by a smaller metallicity spread ($\delta$[Fe/H]$_{\rm 1G}^{\rm MAX}=0.037\pm0.008$).

\begin{figure}
    \centering
    \includegraphics[width=.95\columnwidth]{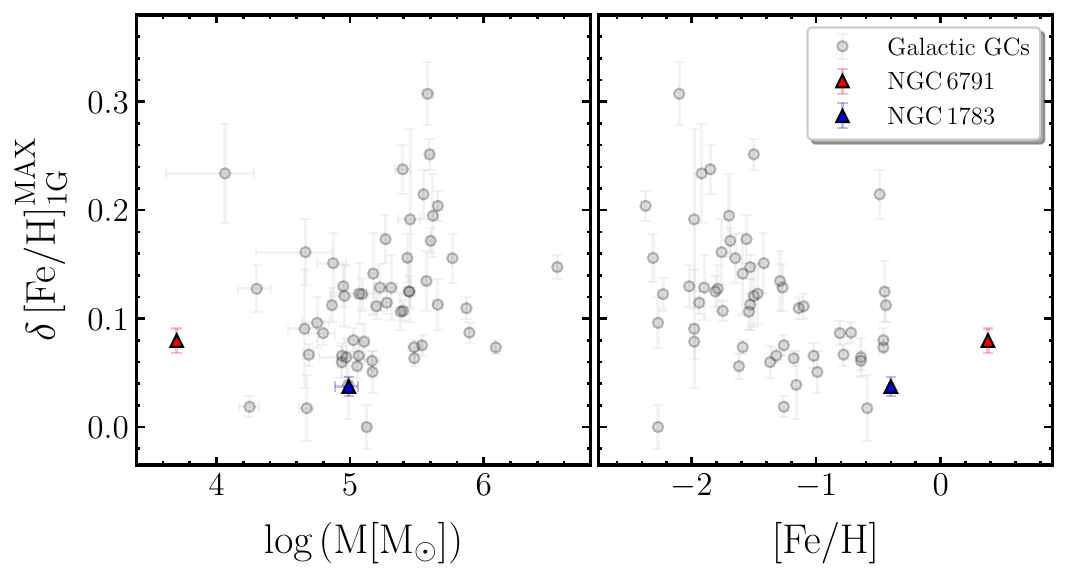}       
    \caption{Maximum iron variations of 1G stars, $\delta$[Fe/H]$_{\rm 1G}^{\rm MAX}$, as a function of the logarithm of the cluster mass (left) and metallicity (right). Gray dots correspond to the 55 Galactic GCs analyzed by \cite{legnardi2022}, whereas NGC\,6791 and NGC\,1783 are marked with red and blue triangles, respectively.}
    \label{fig:SSPs_corrs}
\end{figure}

To compare NGC\,6791 and NGC\,1783 with the 1G stars of Galactic GCs, we plot in the left and right panels of Fig.~\ref{fig:SSPs_corrs} the 1G [Fe/H] variations as a function of the logarithm of the cluster mass \citep[from][]{baumgardt2018} and metallicity \citep[from][2010 update]{harris1996}, respectively. For NGC\,6791 we used the present-day mass and the metallicity computed by \citet{cordoni2023a} and \citet{frinchaboy2013}, respectively, whereas for NGC\,1783 we exploited the values derived from \citet{milone2023b} and \citet{song2021}. 

In both panels gray dots correspond to the 55 Galactic GCs studied by \cite{legnardi2022}, whereas the filled triangles indicate NGC\,6791 (red) and NGC\,1783 (blue). The $\delta$[Fe/H]$_{\rm 1G}^{\rm MAX}$ values of Galactic GCs exhibit a mild correlation/anticorrelation with the cluster mass/metallicity. Both clusters follow the general trend of both relations, despite the large [Fe/H] variation observed among clusters with similar mass or metallicity.

\section{First-generation stars in Galactic globular clusters}
\label{sec:GGCs}
The 1G color broadening is a widespread phenomenon among Galactic GCs. However, the collection of ChMs by \citet[][see Figs.~3-7]{milone2017} and \citet[][see Figs.~12 and 13]{jang2022a} reveal that the $\Delta_{\rm F275W,F814W}$ pseudo-color extension of 1G stars exhibit substantial variations among different clusters. By exploiting the color-metallicity relations by \citet{dotter2008}, \citet{legnardi2022} converted the $\Delta_{\rm F275W, F814W}$ values into [Fe/H] variations for a large sample of 55 Galactic GCs. Similarly to the $\Delta_{\rm F275W, F814W}$ pseudo-color, the resulting values of $\delta$[Fe/H]$_{\rm 1G}^{\rm MAX}$ changes dramatically from one cluster to another, ranging from $\sim 0.00$ dex to $\sim 0.30$ dex \citep[see Fig.~12 of][for more details]{legnardi2022}.

\begin{figure*}
    \centering
    \includegraphics[width=.95\textwidth,trim={1cm 1cm 2.3cm 17.45cm},clip]{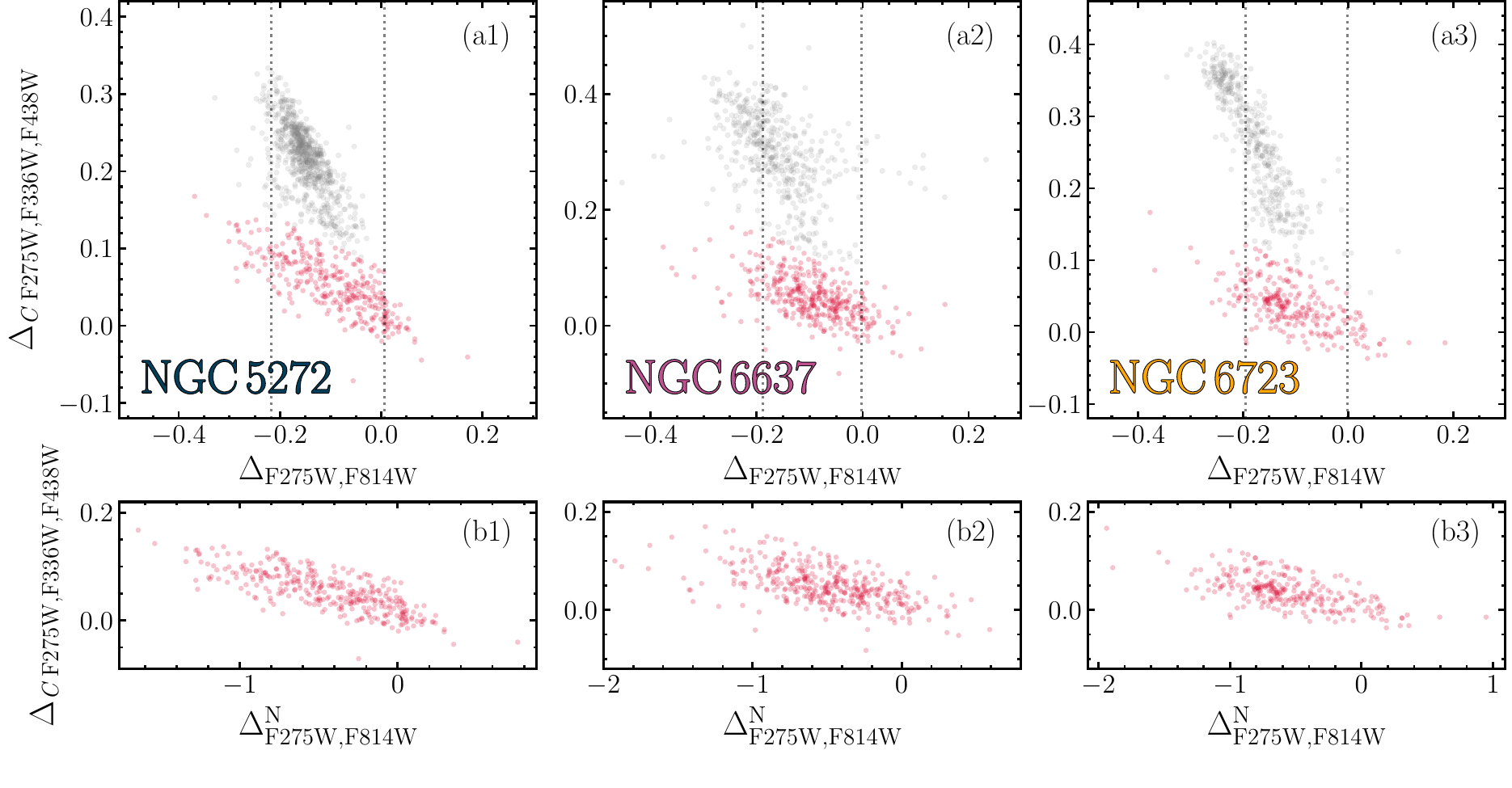}
    \includegraphics[width=.95\textwidth,trim={1cm 1cm 2.7cm 25.15cm},clip]{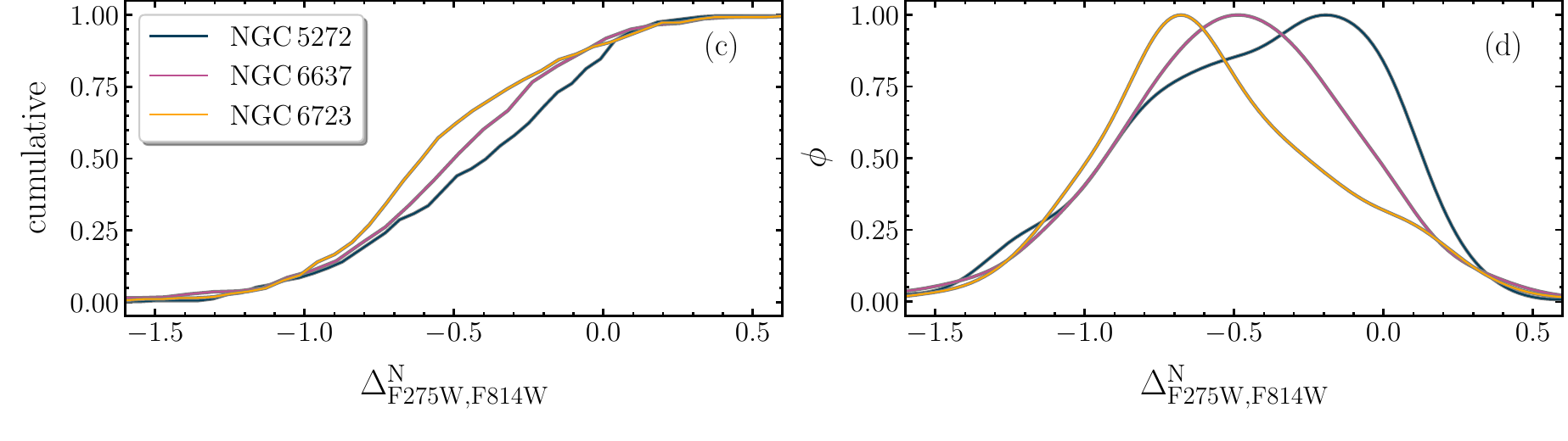}
    \caption{{\it Top panels.} ChMs of NGC\,5272 (a1), NGC\,6637 (a2), and NGC\,6723 (a3) where the 1G stars are colored red. {\it Middle panels.} Zoom of the ChMs shown in the top panels for 1G stars of NGC\,5272 (b1), NGC\,6637 (b2), and NGC\,6723 (b3). Here, we normalize the $\Delta_{\rm F275W,F814W}$ pseudo-color to the width of the 1G sequence. {\it Bottom panels.} Cumulative (c) and kernel-density (d) distributions of the $\Delta_{\rm F275W,F814W}^{\rm N}$ pseudo-color for 1G stars of NGC\,5272 (dark blue), NGC\,6637 (purple), and NGC\,6723 (orange).}
    \label{fig:cumdist}
\end{figure*}

Additionally, we noticed that the ChMs of the 55 GCs studied by \cite{legnardi2022} reveal different distribution of the 1G $\Delta_{\rm F275W,F814W}$ pseudo-color. This fact is illustrated in panels a1-a3 of Fig.~\ref{fig:cumdist}, where we present the ChMs of NGC\,5272, NGC\,6637, and NGC\,6723 that we consider as test cases. The red points represent 1G stars, while the vertical dotted lines indicate the 10th and the 90th percentiles of the 1G $\Delta_{\rm F275W,F814W}$ distribution. To properly compare the pseudo-colors of 1G stars in the different clusters, we normalized the $\Delta_{\rm F275W,F814W}$ pseudo-colors by the distance between the two vertical lines shown in the top panels of Fig.~\ref{fig:cumdist}. Results are illustrated in panels b1-b3 of Fig.~\ref{fig:cumdist}, where we show the $\Delta_{\rm {\it C}\, F275W,F336W,F438W}$ vs.\,$\Delta^{\rm N}_{\rm F275W,F814W}$ diagrams for 1G stars alone.

A visual inspection at the top and middle panels of Fig.~\ref{fig:cumdist} reveals that most 1G stars in NGC\,5272 and NGC\,6273 exhibit low and large values of $\Delta^{\rm N}_{\rm F275W,F814W}$, respectively. Consequently, the $\delta$[Fe/H] distribution of their 1G stars is dominated by metal-poor and metal-rich stars, respectively. NGC\,6637 represents an intermediate case. The differences among the pseudo-color distributions of 1G stars in the ChMs of NGC\,5272, NGC\,6637, and NGC\,6723 are highlighted by the cumulative and the kernel-density distributions plotted in the bottom panels of Fig.~\ref{fig:cumdist}. 

In the following, we further investigate 1G stars in a large sample of 51 Galactic GCs. Specifically, in Sect.~\ref{subsec:Apar} we quantify the $\Delta_{\rm F275W,F814W}$ pseudo-color distribution, while in Sect.~\ref{subsec:corr} we investigate possible relations with the main parameters of the host GC.

\subsection{The color distribution of first-generation stars in 51 Galactic globular clusters}
\label{subsec:Apar}
\begin{figure}
    \centering
    \includegraphics[width=.95\columnwidth]{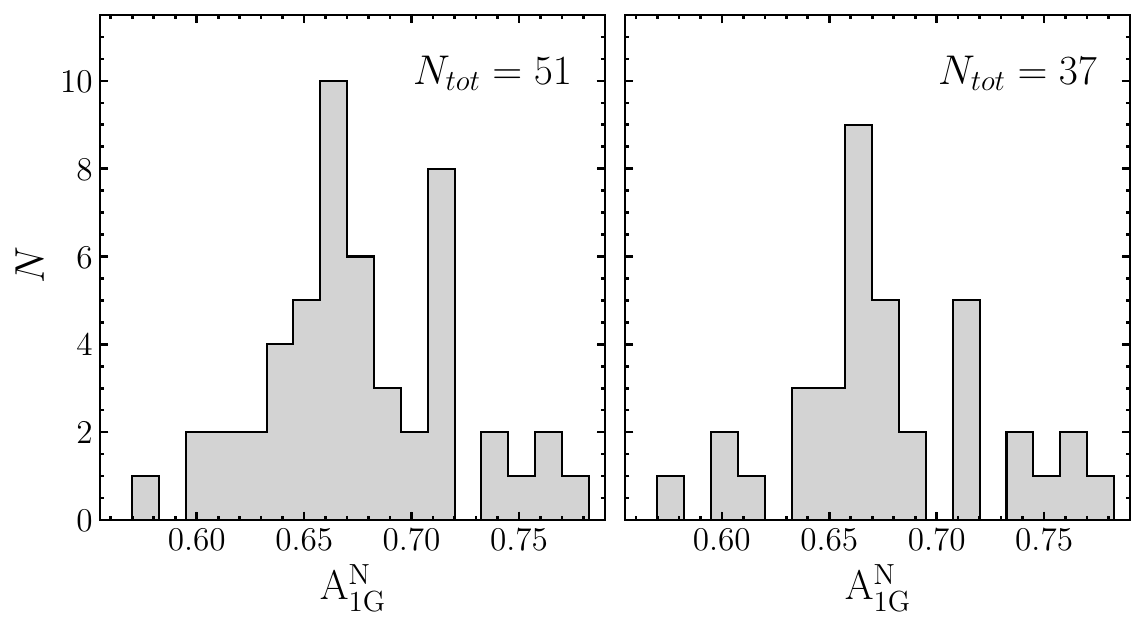}
    \caption{Histograms of the A$^{\rm N}_{\rm 1G}$ parameter for all the 51 GCs (left panel) and the selected 37 GCs with wide 1G sequences and well-separated groups of 1G and 2G stars in the ChM (right panel). See the text for details.}
    \label{fig:A_hist}
\end{figure}

To parameterize the $\Delta^{\rm N}_{\rm F275W,F814W}$ distribution of 1G stars, we calculated the area under the cumulative curve, denoted as A$^{\rm N}_{\rm 1G}$, for 51 Galactic GCs\footnote{We excluded from this analysis the clusters for which the ChM shows no appreciable separation between 1G and 2G stars, namely NGC\,5927, NGC\,6304, NGC\,6388, and NGC\,6441. Additionally, we excluded $\omega$\,Centauri, due to the complexity of its ChM.}. Results are listed in Table~\ref{A_table}. As illustrated in the left panel of Fig.~\ref{fig:A_hist}, where we show the histogram distribution of A$^{\rm N}_{\rm 1G}$, the values of A$^{\rm N}_{\rm 1G}$ range from $\sim 0.6$ to $\sim 0.8$ mag, with large and low values indicating a predominance of stars with small and large $\Delta^{\rm N}_{\rm F275W,F814W}$ values, corresponding to metal-poor and metal-rich stars, respectively. The distribution exhibits a main peak around A$^{\rm N}_{\rm 1G}=0.65$ mag plus a tail towards large values of A$^{\rm N}_{\rm 1G}$. When we limited the analysis to the GCs with wide 1G sequences and where the 1G stars are clearly separated by the 2G in the ChM\footnote{The selected clusters have [Fe/H]$>-0.64$ and 1G pseudo-color extension, $W_{\rm F275W,F814W}^{\rm 1G}$, larger than 0.074 mag.}, we obtained the same qualitative result. This conclusion is supported by the similarity between the A$^{\rm N}_{\rm 1G}$ histogram distributions derived for all clusters and the selected GCs alone (left and right panels of Fig.~\ref{fig:A_hist}, respectively).

\subsection{Relations between the first-generation color distribution and the parameters of the host cluster}
\label{subsec:corr}
To test whether the distribution of metal-rich and metal-poor 1G stars can be influenced by the global parameters of the host GC, in the following we investigate the relation between A$^{\rm N}_{\rm 1G}$ and the main GC parameters. Our analysis includes metallicity ([Fe/H]), reddening ($E(B-V)$), absolute visual magnitude ($M_{\rm V}$), central surface brightness (SB$_{0}$), central stellar density ($\rho_{0}$), ellipticity ($\epsilon$), concentration (c), and specific RR Lyrae density (S$_{\rm RR}$) from the 2010 version of the \cite{harris1996} catalog. Additionally, we used various parameters from \cite{baumgardt2018}, including total cluster mass (M), mass-to-light ratio in the {\it V} band (M/L), core density ($\rho_{c}$), half-mass-radius density ($\rho_{hm}$), half-mass relaxation time ($T_{RH}$), slope of mass function (MF slope), mass fraction of the remnants ($F_{\rm remn}$), central velocity dispersion ($\sigma_{0}$), central escape velocity ($v_{esc}$), and mass segregation parameter inside the core ($\eta_{c}$) and half-mass–radius ($\eta_{hm}$). 

GC ages have been taken from \citet[][hereafter \citetalias{marinfranch2009}]{marinfranch2009}, \citet[][hereafter \citetalias{dotter2010}]{dotter2010}, \citet[][hereafter \citetalias{vandenberg2013}]{vandenberg2013}, and \citet[][hereafter \citetalias{tailo2020}]{tailo2020}, whereas the fraction of binaries in GCs comes from \cite{milone2012a}, as measured within the core (f$_{\rm bin,c}$), between the core and the half-mass radius (f$_{\rm bin,hm}$), and beyond the half-mass radius (f$_{\rm bin,ohm}$).  

Moreover, we took advantage of the initial mass of the cluster (M$_{\rm i}$), the distance from the Galactic Center (R$_{\rm GC}$), the mean heliocentric velocity (RV), the apogalacticon and the perigalaticon radius (R$_{\rm apog}$ and R$_{\rm perig}$) from \cite{baumgardt2019}. HB parameters, such as the 1G and extreme-2G mass loss ($\mu_{\rm 1G}$ and $\mu_{\rm 2Ge}$), the difference between the two ($\delta \mu_{\rm e}$), and the 1G and 2Ge average HB mass ($\overline{M}_{\rm 1G}^{\rm HB}$ and $\overline{M}_{\rm 2Ge}^{\rm HB}$), have been computed by \cite{tailo2020}, while the RGB width in $m_{\rm F275W}-m_{\rm F814W}$ (W$_{\rm F275W}$) and in $C_{\rm F275W,F336W,F438W}$ (W$_{\rm C,F275W}$), the RGB width of the 1G in $m_{\rm F275W}-m_{\rm F814W}$ (W$_{\rm F275W}^{\rm 1G}$), the RGB width in $C_{\rm F275W,F336W,F438W}$ without [Fe/H] dependency ($\Delta$W$_{\rm C,F275W}$), and the fraction of 1G stars (N$_{\rm 1G}$/N$_{\rm T}$) have been derived from \cite{milone2017}.

We used the mean and maximum internal helium enhancement ($\delta$Y$_{\rm 2G,1G}$ and $\delta$Y$_{\rm max}$) together with the internal helium variation within the 1G ($\delta$Y$_{\rm 1G}$) from \cite{milone2018a}. The remaining parameters, namely the HB ratio (HBR), the F606W$-$F814W color distance of the red HB from the RGB (L$_{1}$), the HB F606W$-$F814W color extension (L$_{2}$), the RGB width in $C_{\rm F336W,F438W,F814W}$ with and without [Fe/H] dependency (W$_{\rm C,F336W}$ and $\Delta$W$_{\rm C,F336W}$), the internal metallicity variations within the 1G ($\delta$[Fe/H]$_{\rm 1G}^{\rm MAX}$), the mean differential reddening variation ($\sigma_{\rm A_{F814W}}$) have been taken from \citet[][HBR]{mackey2005}, \citet[][L$_{1}$ and L$_{2}$]{milone2014}, \citet[][W$_{\rm C,F336W}$ and $\Delta$W$_{\rm C,F336W}$]{lagioia2019}, \citet[][$\delta${[Fe/H]}$_{\rm 1G}^{\rm MAX}$]{legnardi2022}, and \citet[][$\sigma_{\rm A_{F814W}}$]{legnardi2023}, respectively. Finally, we also included the mass of 1G stars (M$_{\rm 1G}$) and the dynamical age of the cluster (Age/$T_{RH}$), computed by using the fraction of 1G stars \citep{milone2017} and the half-mass relaxation time \citep{baumgardt2018}, respectively.

To estimate the statistical correlation between A$^{\rm N}_{\rm 1G}$ and the parameters listed above, we calculated the Spearman’s rank coefficient (R$_{\rm S}$) and the corresponding p-value, which indicates the significance of the correlation. Results are listed in Table~\ref{correlations_table}, where we provide R$_{\rm S}$, the p-value, and the number of freedom degrees for each couple of quantities. Intriguingly, A$^{\rm N}_{\rm 1G}$ exhibits no significant correlation with the considered parameters. We repeated this analysis after dividing the investigated targets in M\,3- and M\,13-like clusters according to the morphology of their HBs \footnote{We defined M\,3- and M\,13-like GCs objects with L$_{1}\le0.35$ and L$_{1}>0.35$, respectively \citep[see][for details]{legnardi2022}.}. We found no strong correlation between A$^{\rm N}_{\rm 1G}$ and the global GC parameters neither for the M\,3-like nor the M\,13-like GC groups. While the maximum iron variation within 1G stars correlates with GC mass \citep{legnardi2022}, the result of this analysis implies that the relative abundance of metal-rich and metal-poor 1G stars within a cluster is not influenced by the intrinsic properties of the host GC.

\begin{table*}
    \centering
    \caption{Area under the cumulative curve of the 1G $\Delta^{\rm N}_{\rm F275W,F814W}$ distribution (A$^{\rm N}_{\rm 1G}$) for 51 Galactic GCs.}
    \begingroup
    \setlength{\tabcolsep}{10pt} 
    \renewcommand{\arraystretch}{1.} 
    \begin{tabular}{cccccc}
    \hline \\[-.3cm]
         Cluster ID & A$^{\rm N}_{\rm 1G}$ & Cluster ID & A$^{\rm N}_{\rm 1G}$ & Cluster ID & A$^{\rm N}_{\rm 1G}$ \\[.1cm]
         \hline \\[-.3cm]
         NGC\,0104 & $0.718\pm0.025$ & NGC\,5986 & $0.716\pm0.031$ & NGC\,6584 &  $0.764\pm0.032$ \\  
         NGC\,0288 & $0.666\pm0.040$ & NGC\,6093 & $0.645\pm0.028$ & NGC\,6624 &  $0.653\pm0.036$ \\
         NGC\,0362 & $0.687\pm0.029$ & NGC\,6101 & $0.699\pm0.037$ & NGC\,6637 &  $0.666\pm0.024$ \\
         NGC\,1261 & $0.759\pm0.025$ & NGC\,6121 & $0.714\pm0.097$ & NGC\,6652 &  $0.639\pm0.044$ \\
         NGC\,1851 & $0.635\pm0.030$ & NGC\,6144 & $0.713\pm0.059$ & NGC\,6656 &  $0.660\pm0.045$ \\
         NGC\,2298 & $0.606\pm0.068$ & NGC\,6171 & $0.668\pm0.047$ & NGC\,6681 &  $0.702\pm0.039$ \\
         NGC\,2808 & $0.736\pm0.017$ & NGC\,6205 & $0.660\pm0.032$ & NGC\,6717 &  $0.661\pm0.067$ \\
         NGC\,3201 & $0.609\pm0.058$ & NGC\,6218 & $0.651\pm0.040$ & NGC\,6723 &  $0.746\pm0.027$ \\
         NGC\,4590 & $0.713\pm0.041$ & NGC\,6254 & $0.675\pm0.037$ & NGC\,6752 &  $0.668\pm0.047$ \\
         NGC\,4833 & $0.655\pm0.037$ & NGC\,6341 & $0.676\pm0.028$ & NGC\,6779 &  $0.670\pm0.034$ \\
         NGC\,5024 & $0.665\pm0.025$ & NGC\,6352 & $0.596\pm0.046$ & NGC\,6809 &  $0.580\pm0.061$ \\
         NGC\,5053 & $0.680\pm0.100$ & NGC\,6362 & $0.740\pm0.037$ & NGC\,6838 &  $0.719\pm0.045$ \\
         NGC\,5272 & $0.646\pm0.028$ & NGC\,6366 & $0.626\pm0.081$ & NGC\,6934 &  $0.717\pm0.031$ \\
         NGC\,5286 & $0.715\pm0.022$ & NGC\,6397 & $0.612\pm0.081$ & NGC\,6981 &  $0.770\pm0.030$ \\
         NGC\,5466 & $0.630\pm0.074$ & NGC\,6496 & $0.641\pm0.039$ & NGC\,7078 &  $0.679\pm0.020$ \\
         NGC\,5897 & $0.670\pm0.049$ & NGC\,6535 & $0.642\pm0.097$ & NGC\,7089 &  $0.669\pm0.032$ \\
         NGC\,5904 & $0.683\pm0.031$ & NGC\,6541 & $0.668\pm0.029$ & NGC\,7099 &  $0.693\pm0.050$ \\[.1cm]
         \hline \hline
    \end{tabular}
    \endgroup     
    \label{A_table}
\end{table*}

\begin{table*}
    \centering
    \caption{Relation between A$^{\rm N}_{\rm 1G}$ and the main GC parameters. For each couple of quantities, we list the Spearman's rank coefficient, the corresponding p-value, and the number of freedom degrees.}
    \begingroup
    \setlength{\tabcolsep}{12pt} 
    \renewcommand{\arraystretch}{1.2} 
    \begin{tabular}{cccccc}
    \hline \\[-.3cm]
        Parameter &  A$^{\rm N}_{\rm 1G}$ & Parameter & A$^{\rm N}_{\rm 1G}$ & Parameter & A$^{\rm N}_{\rm 1G}$   \\[.15cm]
         \hline \\[-.3cm] 
          Age \citepalias{marinfranch2009} & $-0.03,0.840,48$ & W$_{\rm F275W}$ & $-0.01,0.966,49$ & $\mu_{\rm 1G}$ & $-0.12,0.435,41$\\
          Age \citepalias{dotter2010} & $-0.23,0.098,49$ & W$_{\rm C,F275W}$ & $0.04,0.783,49$ & $\mu_{\rm 2Ge}$ & $0.05,0.793,32$\\
          Age \citepalias{vandenberg2013} & $-0.05,0.754,46$ & $\Delta$W$_{\rm C,F275W}$ & $0.04,0.783,49$ & $\overline{M}_{\rm 1G}^{\rm HB}$ & $0.06,0.683,41$ \\ 
          Age \citepalias{tailo2020} & $-0.13,0.416,41$ & W$_{\rm C,F336W}$ & $0.16,0.276,49$ & $\overline{M}_{\rm 2Ge}^{\rm HB}$ & $-0.01,0.934,32$ \\
          $\delta$Y$_{\rm 2G,1G}$ & $-0.09,0.522,48$ & $\Delta$W$_{\rm C,F336W}$ & $0.21,0.147,49$ & $\delta \mu_{\rm e}$ & $0.02,0.924,32$ \\
          $\delta$Y$_{\rm max}$ & $0.10,0.500,48$ & W$_{\rm F275W}^{\rm 1G}$ & $0.04,0.794,49$ & L$_{1}$ & $-0.11,0.433,48$ \\
          $\delta$Y$_{\rm 1G}$ & $0.13,0.349,49$ & $E(B-V)$ & $-0.12,0.405,49$ & L$_{2}$ & $0.34,0.015,48$ \\
          $\delta$[Fe/H]$_{\rm 1G}^{\rm MAX}$ & $-0.03,0.826,49$ & $\sigma_{\rm A_{F814W}}$ & $-0.26,0.062,49$ & S$_{\rm RR}$ & $0.24,0.087,49$ \\
          f$_{\rm bin,c}$ & $-0.19,0.276,32$ & SB$_{0}$ & $-0.10,0.470,49$ & N$_{\rm 1G}$/N$_{\rm T}$ & $-0.05,0.748,49$  \\
          f$_{\rm bin,hm}$ & $-0.18,0.244,43$ & c & $0.02,0.873,49$ & RV & $0.18,0.281,35$  \\
          f$_{\rm bin,ohm}$ & $0.10,0.544,37$ & $\epsilon$ & $-0.11,0.446,49$ & HBR & $-0.12,0.401,48$  \\
          $\log(\rm M/M_{\odot})$ & $0.26,0.071,49$ & $\sigma_{0}$ & $0.15,0.289,49$ & $\rho_{0}$ & $0.01,0.918,49$ \\
          $\log(\rm M_{i}/M_{\odot})$ & $0.12,0.405,49$ & $v_{esc}$ & $0.13,0.368,49$ & $\log(\rho_{c})$ & $-0.07,0.643,49$ \\
          $\log(\rm M_{\rm 1G}/M_{\odot})$ & $0.20,0.169,49$ & MF slope & $-0.17,0.238,49$ & $\log(\rho_{\rm hm})$ & $0.05,0.712,49$ \\
          M/L & $0.06,0.690,49$ & $\eta_{c}$ & $0.00,0.985,49$ & $\log(T_{\rm RH})$ & $0.25,0.083,49$ \\
          $M_{\rm V}$ & $-0.22,0.117,49$ & $\eta_{hm}$ & $0.12,0.415,49$ & $\log({\rm Age}/T_{RH})$ & $-0.26,0.062,49$ \\
          R$_{\rm apog}$ & $0.18,0.216,49$ & [Fe/H] & $0.06,0.659,49$ & F$_{\rm remn}$ & $-0.12,0.407,49$ \\
          R$_{\rm GC}$ & $0.18,0.218,49$ & R$_{\rm perig}$ & $0.00,0.989,49$ & &  \\[.1cm]
         \hline \hline
    \end{tabular}
    \endgroup
    \label{correlations_table}
\end{table*}

\section{Summary and Discussions}
\label{sec:concl}
It is now widely accepted that the 1G stars within GCs commonly display variations in metallicity from star to star, leading to prolonged sequences in the ChMs of MS and RGB stars \citep[e.g.,][]{legnardi2022, lardo2022, marino2019a, marino2019b, marino2023, tailo2019}. While the metal content of 1G stars has been thoroughly examined in the central regions of an extensive sample encompassing 55 GCs, several unresolved questions persist. As an example, the radial behavior of the metallicity variations within the GCs is still unexplored. Moreover, the sample of studied clusters is entirely composed of old Galactic GCs with multiple stellar populations and ages of more than $\sim$12 Gyr \citep[e.g.,][]{dotter2010, tailo2020}. Hence, we do not know whether metallicity variations are a peculiarity of old Milky Way GCs or are also present in Galactic or extragalactic star clusters younger than 12 Gyr. Finally, the relations between the metallicity distributions of 1G stars and the main parameters of the host clusters are poorly studied. In this work, we investigated these open issues by providing an in-depth analysis of 1G stars in different environments, including Galactic and extragalactic star clusters. 
 
Our first analysis was focused on 47\,Tucanae. This dynamically young Galactic GC is an ideal target to investigate the radial behavior of 1G stars with different metallicities. Indeed, the fact that its 2G stars are significantly more centrally concentrated than the 1G, suggests that the stellar populations of 47\,Tucanae retain a memory of their initial distribution during the cluster formation. Specifically, we used HST and JWST data to study 1G stars in the cluster core and three external fields, located at $\sim 7$, $\sim 8.5$, and $\sim 11$ arcmin from the cluster center, respectively. The results can be summarized as follows:
\begin{itemize}
    \item We derived the ChMs of RGB stars in the central field and of M-dwarf stars in the external fields. We detected a well-elongated 1G sequence in each ChM, which is not consistent with a chemically homogeneous stellar population.
    \item We took advantage of the pseudo-color elongation along the x-axis of the ChM to infer the [Fe/H] distribution of 1G stars. All the distributions span similar intervals, included between $\delta$[Fe/H]$_{\rm 1G} \sim -0.15$ dex and $\sim 0.05$ dex, and display a major peak at $\delta$[Fe/H]$_{\rm 1G} \sim -0.07$ dex. The [Fe/H] distribution of the cluster center shows an additional peak centered at $\delta$[Fe/H]$_{\rm 1G} \sim -0.02$, which gradually disappears moving outwards. This fact suggests that 1G stars richer in metals are more centrally concentrated than the metal-poor ones. 
    \item We identified two main groups of metal-rich and metal-poor 1G stars in the four studied fields and used stellar proper motions to investigate their internal kinematics. We found that both groups of 1G stars share similar radial and tangential velocity dispersion profiles and are characterized by isotropic motions.
\end{itemize}

\cite{mckenzie2021} used hydrodynamical simulations to reproduce a giant molecular cloud forming GCs in a high-redshift dwarf galaxy. The simulated clusters exhibit star-to-star metallicity variations of $\sim 0.1$ dex,  which are consistent with the [Fe/H] spread that we observe among 1G stars in 47\,Tucanae \citep[see also][]{legnardi2022, marino2023}. Such metallicity variations are due to the merging of gas clumps and self-enrichment processes. Intriguingly, the evidence of a small increase in the fraction of metal-rich stars in 47\,Tucanae is consistent with the properties of most forming GCs simulated by \cite{mckenzie2021}, where the metal-rich stars are more centrally concentrated than the metal-poor ones.

We notice that while the metallicity distribution of the central field significantly differs from those of the external fields, we detect a less pronounced metallicity gradient only when we consider the three external fields alone. In particular, the average metallicity variations among the 1G stars of these fields differ by $\sim 3 \sigma$ only. Since the results for the central field are obtained from RGB stars, whereas the conclusions of the external regions are derived from M dwarfs, an alternative explanation is that the observed metallicity-distribution differences in the central and external fields are not due to a radial gradient but reflect different mass functions for the groups of stars with different metallicities.

Recent works have suggested that the properties of stellar populations with different light-element abundances do not significantly depend on stellar mass. This conclusion is based on the evidence that stars with different masses in various GCs, including 47\,Tucanae, share the same [O/Fe] abundances and that their 1G and 2G stars have similar mass functions \citep{milone2014b, milone2019a, dondoglio2022, marino2024a}. The potential influence of stellar mass on the observed variances in the metallicity distributions of 1G stars would not alter these conclusions. It is worth noting that the distinct mass functions among groups of metal-poor and metal-rich stars within the 1G do not significantly affect the mass functions and the oxygen abundances of 1G and 2G stars\footnote{In the scenarios based on multiple generations, the 2G stars formed from material that is polluted (in light elements) by the ejecta of more massive 1G stars. A mass-function difference between the 1G metal-rich and metal-poor stars would affect the amount of metal-rich and metal-poor polluters. However, due to the small metallicity difference between these two groups of stars, the chemical composition (i.e., the abundances of He, C, N, O, Mg, and Al) of their ejecta is nearly identical. In the scenarios based on accretion, all GC stars formed in the same star-formation episode. The chemical composition of the 2G stars is due to accreted polluted material in the early stages of star formation. Also in this case, we do not expect that the mass functions of the groups of metal-rich and metal-poor 1G stars would affect the accretion process.}.

To investigate whether the 1G color extension phenomenon occurs only in clusters hosting multiple stellar populations or not, we analyzed HST multi-band photometry of two simple-population star clusters, namely the Galactic open cluster NGC\,6791 and the LMC cluster NGC\,1783. The main outcomes of this second analysis include:
\begin{itemize}
    \item We derived the standard $\Delta_{\rm {\it C}\,F275W, F336W, F438W}$ vs.\,$\Delta_{\rm F275W,F814W}$ ChM for RGB stars in NGC\,6791 and NGC\,1783. The ChMs of both clusters exhibit one sequence of stars alone, which resembles the 1G sequence of GCs with multiple populations.  This fact confirms that both clusters are consistent with simple populations with homogeneous nitrogen abundances. The evidence that NGC\,6791 hosts stars with different metallicities is confirmed by wide-field ground-based photometry in the U, B, V, and I bands. 
    \item The F275W$-$F814W color extension of both NGC\,6791 and NGC\,1783 is much wider than the color spread due to the observational errors alone. Hence, we concluded that extended sequences in the ChM are not prerogatives of 1G stars in GCs but are also present in simple-population clusters.
    \item We estimated the maximum [Fe/H] variations among the stars of NGC\,6791 and NGC\,1783, by assuming that iron spreads are the main drivers of the color extension observed on the ChM. We found that NGC 6791 exhibits a moderate [Fe/H] variation ($\delta$[Fe/H]$_{\rm 1G}^{\rm MAX}=0.080\pm0.011$ dex), whereas NGC\,1783 shows a much smaller iron spread ($\delta$[Fe/H]$_{\rm 1G}^{\rm MAX}=0.037\pm0.008$ dex). Such metallicity variations are comparable to those observed in metal-rich GCs with similar masses.
\end{itemize}

Based on a sample of 51 ChMs studied by \citet{legnardi2022}, we noticed that the $\Delta_{\rm F275W, F814W}$ pseudo-color distribution significantly changes from one cluster to another. In some GCs, including NGC\,2808 and NGC\,6723, most 1G stars have small $\Delta_{\rm F275W, F814W}$ values. Hence, the $\delta$[Fe/H] distribution of their 1G stars is dominated by metal-poor stars. Conversely, other clusters, such as NGC\,5272 and NGC\,3201, host larger fractions of metal-rich stars with large values of $\Delta_{\rm F275W, F814W}$. To parameterize the color distribution of 1G stars, we computed the area under the cumulative curve of the normalized $\Delta_{\rm F275W,F814W}$ distribution, A$^{\rm N}_{\rm 1G}$. Intriguingly, the values of A$^{\rm N}_{\rm 1G}$ exhibit no significant correlation with any of the 53 analyzed parameters of the host cluster. This fact could indicate that the relative numbers of metal-rich and metal-poor 1G stars do not depend on the properties of the host GC.

\begin{acknowledgements}
We thank the anonymous referee for various suggestions that improved the quality of the manuscript. This work has been funded by the European Union – NextGenerationEU RRF M4C2 1.1 (PRIN 2022 2022MMEB9W: "Understanding the formation of globular clusters with their multiple stellar generations", CUP C53D23001200006), from INAF Research GTO-Grant Normal RSN2-1.05.12.05.10 -  (ref. Anna F. Marino) of the "Bando INAF per il Finanziamento della Ricerca Fondamentale 2022", and from the European Union’s Horizon 2020 research and innovation program under the Marie Skłodowska-Curie Grant Agreement No. 101034319 and the European Union – NextGenerationEU (beneficiary: T. Ziliotto).
\end{acknowledgements}

\bibliographystyle{aa}
\bibliography{aanda}
\end{document}